\documentclass[12pt,aps,prd,twocolumn,nofootinbib,reprint,longbibliography,superscriptaddress]{revtex4-1}
\pdfoutput=1 
\usepackage{amsmath, amssymb, amsfonts, mathrsfs, slashed, bm, graphicx, xcolor, multirow}
\usepackage[colorlinks=true, linkcolor = blue, urlcolor = blue, citecolor = blue, anchorcolor = blue]{hyperref}

\begin{document}
\pagestyle{plain}

\title{Exploring the Electromagnetically Interacting Dark Matter at the International Linear Collider}

\author{Manish Kumar Sharma}
\email{p20190006@goa.bits-pilani.ac.in}
\affiliation{Department of Physics, Birla Institute of Technology and Science Pilani, K K Birla Goa Campus, NH-17B, Zuarinagar, Goa 403726, India}

\author{Saumyen Kundu}
\email{p20170022@goa.bits-pilani.ac.in}
\affiliation{Department of Physics, Birla Institute of Technology and Science Pilani, K K Birla Goa Campus, NH-17B, Zuarinagar, Goa 403726, India}

\author{Prasanta Kumar Das}
\email{pdas@goa.bits-pilani.ac.in}
\affiliation{Department of Physics, Birla Institute of Technology and Science Pilani, K K Birla Goa Campus, NH-17B, Zuarinagar, Goa 403726, India}

\begin{abstract}
  Dark Matter being electrically neutral does not participate in electromagnetic interactions at leading order. However, we discuss here fermionic dark matter (DM) with permanent magnetic and electric dipole moment that interacts electromagnetically with photon at loop-level through a dimension-5 operator. We discuss the search prospect of the dark matter at the proposed International Linear Collider (ILC) and constrain the parameter space in the plane of the DM mass and the cutoff scale $\Lambda$. At the 500 GeV ILC with $4$ ab$^{-1}$ of integrated luminosity we probed the mono-photon channel and utilizing the advantages of beam polarization we obtained an upper bound on the cutoff scale that reaches up to $\Lambda = 3.72$ TeV. 
\end{abstract}
\maketitle

\section{Introduction}
The existence of dark matter (DM) is well established from numerous cosmological and astrophysical observations at different length scales of the Universe. These observations such as the flatness of galactic rotation curves, velocities of stars in dwarf galaxies, velocities of galaxies in clusters, hot gas in galaxy clusters, collisions of galaxy clusters, and gravitational lensing are all results of gravitational interaction with the visible sector, and collisionless property~\cite{Sofue:2000jx,Chang:2020rem,Zwicky:1937zza,Clowe:2006eq,Massey:2010hh}. Experiments like WMAP~\cite{Bennett_2013} and PLANCK~\cite{Planck:2018vyg} analyzing the anisotropy of the cosmic microwave background (CMB) also reaffirm the existence of DM with quantified results that $\sim27\%$ of the total energy budget of the Universe is contributed by it. Despite all these observations, the particle nature of the DM is still elusive. In a bid to understand this, numerous theoretical models have been put forward and several experimental efforts are being carried out and have been proposed. 
Of these theoretical models, the WIMP paradigm is the most explored because of the possibility of a weak scale mass and coupling which motivates different BSM models. However, no conclusive evidence of WIMP DM observation has been achieved. In fact, recent bounds from direct detection experiments both from DM-electron as well as DM-nucleon scattering are pushing the DM parameter space down to the neutrino floor~\cite{Billard:2013qya} or the neutrino fog~\cite{PhysRevLett.127.251802} region. 

So the community is looking beyond the WIMP paradigm, considering alternate models based on the production mechanism (e.g., freeze-in DM)~\cite{Bernal:2017kxu}, or interaction with the visible sector (e.g. strongly interacting DM)~\cite{Emken:2019tni}, or interaction with the dark sector (e.g. self-interacting DM)~\cite{Mavromatos:2017lfs}, or with electric charge~\cite{Stebbins:2019xjr}, or with fractional charge (e.g. millicharged DM)~\cite{Davidson:2000hf} and so on. We have considered here fermionic DM that have zero electric charge or have a very tiny charge so that they practically do not interact with the photons (hence `dark') at leading order, but have non-zero intrinsic magnetic or electric or both the dipole moments. Due to this intrinsic dipole moment, they interact with the photons at the 1-loop order or higher~\cite{Hambye:2021xvd}. This class of DMs, termed as \emph{dipole dark matter} (DDM), has been studied extensively in the literature in different cosmological and astrophysical context~\cite{Sigurdson:2004zp, Mahmoudi:2018lll, BARGER201174, Geytenbeek:2016nfg, Binh:2020xtf, Aranda:2015jis, Hambye:2021xvd, Lopes:2013xua, PhysRevD.80.036009, Chang:2019xva}. 

The DDMs are motivated by the possibility that the electromagnetic interactions may improve the direct detection sensitivity~\cite{Barger:2010gv, Fitzpatrick:2010br, Banks:2010eh, DelNobile:2012tx, Sigurdson:2004zp, PhysRevD.80.036009}. Models of DM as Dirac, Majorana, or pseudo-Dirac fermions with small mass-splitting are studied in various literature that consider different electromagnetic form factors for the fermionic DM candidates. In ref.~\cite{Kopp:2014tsa} Dirac and Majorana DDMs with a leptophilic nature are discussed in the direct detection context giving bounds on anapole and dipole moments. DDMs are also studied in the context of energy transport in the solar interior~\cite{Geytenbeek:2016nfg}, helioseismological study~\cite{Lopes:2013xua}, scattering from the Sun~\cite{Chatterjee:2022gbo}, $\gamma$-ray searches~\cite{DelNobile:2012tx}, CMB searches~\cite{Mahmoudi:2018lll}, extra dimension~\cite{Flacke:2006ut}, collider search~\cite{Kadota:2014mea, Dienes:2023uve, Kling:2022ykt, Fortin:2011hv} and so on. In this work, we extend the collider probe of DDM at the proposed International Linear Collider (ILC) as a complementary study to the direct and indirect detection searches in an effort to present a detailed collider analysis presenting an updated bound on the DDM parameter space. 

The most promising channel for searching the weak-scale DM comprises of a characteristic mono-$X$ signature, where the large missing transverse momentum carried away by the DM pair is balanced by a visible sector particle $X$ (which can be either a photon, jet, $W$, $Z$ or Higgs, depending on the model) emitted from an initial, intermediate or final state (see Refs.~\cite{Kahlhoefer:2017dnp, Penning:2017tmb} for reviews). However, in the busy environment of the Large Hadron Collider (LHC), the mono-$X$ search does not provide satisfactory results. A workaround can be looking at the far-forward detectors such as FRASER (ForwArd Search ExpeRiment )~\cite{Dienes:2023uve}, or FLArE (Forward Liquid Argon Experiment)~\cite{Kling:2022ykt}. Another good alternative can be the lepton colliders like the International Linear Collider (ILC), Compact Linear Collider (CLIC), or the muon collider because of their clean environment. ILC with the advantage of polarizing both the incoming helps to minimize the SM background contribution and naturally presents itself as a very attractive choice to probe the DDM, particularly through the mono-photon channel~\cite{Chae:2012bq, Kadota:2014mea, Habermehl:2020njb, Kalinowski:2021tyr, Kundu:2021cmo}.
 \par
Here we assume that the DM is a Dirac fermion and it has magnetic dipole moment ($\mu_\chi$) and electric dipole moment ($d_\chi$)  form factors through which it couples to photon via dimension-$5$ operator. The relevant degrees of freedom in our analysis are the DM mass ($m_\chi$) and an effective cutoff scale $\Lambda$ which determines the dark matter-photon coupling i.e. the strength of the dimension-$5$ effective operator \cite{DelNobile:2012tx,Kadota:2014mea}. We make a comprehensive study of the mono-photon signature of dark matter ($\chi$) at future $e^+e^-$ collider (ILC: International Linear Collider) with the center-of-mass energy $\sqrt s=500$ GeV and compare our results with others.
The rest of the paper is organized as follows. In \autoref{sec:2}, we briefly describe the model Lagrangian for the DM-photon in the fermion DM scenario. \autoref{sec:3} discusses the existing constraints from different direct detection experiments. In \autoref{sec:4}, we present our cut-based analysis for the mono-photon$+\slashed{E}_T$ signal and background, both with and without beam polarization. Our summary and conclusion are given in \autoref{sec:4}.

\section{The Model: Probing Dark Matter at ILC} \label{sec:2}
Dark Matter, due to its electrically neutral nature, does not participate in electromagnetic interaction at least in leading order. However, being a fermion they might have interactions at loop-level through its electromagnetic moments.  A typical interaction may lead to an amplitude as 
\begin{equation}
    i\mathcal{M}=i\overline{u}({\bf{p_2}}) \mathcal{O}^\mu(l,q) u({\bf{p_1}}) \epsilon_\mu 
\end{equation}
where, $\mathcal{O}^\mu(l,q)$, the vertex matrix element which acts on the spinors, is a Lorentz vector. Here $l = p_2 + p_1$ and $q = p_2 - p_1$. The most general Lorentz invariant form of $\mathcal{O}^\mu(l,q)$ can be given as~\cite{Nowakowski:2004cv,Giunti:2014ixa,Hambye:2021xvd}
\begin{eqnarray}
    \mathcal{O}^\mu(l,q) = F_Q(q^2)\gamma^\mu &+& F_M(q^2)i\sigma^{\mu\nu} q_\nu + F_E(q^2)\sigma^{\mu\nu}\gamma^5 q_\nu \nonumber \\
    &+& F_A(q^2)(q^2\gamma^\mu - q^\mu\slashed{q})\gamma^5,
\end{eqnarray}
where $q(=p_2-p_1)$ is the photon momentum. For interactions with a real photon ($q^2\to 0$), we have 
\begin{eqnarray}
    F_Q(0)=0,\; F_M(0)= \mu_\chi\text{ (say)},\; F_E(0)&=& d_\chi\text{ (say)},\nonumber \\
    F_A(0)&=& a_\chi\text{ (say)}
\end{eqnarray}
where the form factors are identified, respectively, as the electric charge, magnetic dipole moment, electric dipole moment, and anapole of $\chi$~\cite{Giunti:2014ixa}. We consider here only the magnetic and electric dipole moment interactions as the anapole moment interactions are suppressed being proportional to $q^2$ in the low-$q^2$ limit and are canceled by $\mathcal{O}(q^{-2})$ term of the photon propagator~\cite{Hambye:2021xvd}. 

In this paper, we have considered a dark matter (DM) model where the DM interacts with the photon field through its magnetic and electric dipole moments. Since only Dirac fermions can have permanent dipole moments, our DM particle ($\chi$) is a Dirac fermion, and its interaction with the photon is given by \cite{DelNobile:2012tx,Kadota:2014mea}
\begin{eqnarray}
    \mathcal{L} &=& -\frac{1}{2}\overline{\chi}\left(\mu_{\chi}+i \gamma_5d_{\chi}\right) \sigma_{\mu\nu} \chi F^{\mu\nu} \\ 
     &=& -\left(\frac{1}{\Lambda}\right) \frac{1}{2}\overline{\chi}\left(\mathcal{C}_{M}+i \gamma_5 \mathcal{C}_E\right) \sigma_{\mu\nu} \chi F^{\mu\nu} 
\end{eqnarray}

where $F^{\mu\nu} = \partial^\mu A^\nu - \partial^\nu A^\mu$, is the electromagnetic field strength tensor and $\sigma^{\mu\nu} = \frac{i}{2}[\sigma^\nu, \sigma^\nu]$, the spin tensor. $(\mu_{\chi})$ and $(d_{\chi})$ are magnetic and electric dipole moments respectively and $\Lambda = \mathcal{C}_{M}/\mu_\chi = \mathcal{C}_{E}/d_\chi$ represents the cutoff scale of the effective theory, and $\mathcal{C}_{M}, \mathcal{C}_{E}$ are dimensionless parameters which can take the value $1$ or $0$. We have investigated three different scenarios of dark matter having (i) both Electric and Magnetic, (ii) only Magnetic, and (iii) only Electric dipole moments by considering the values of $(\mathcal{C}_{M},\mathcal{C}_{E})$ as $(1,1),(1,0),(0,1)$ respectively. 
\begin{figure}[t!]
    \centering
    \includegraphics[width=0.95\linewidth]{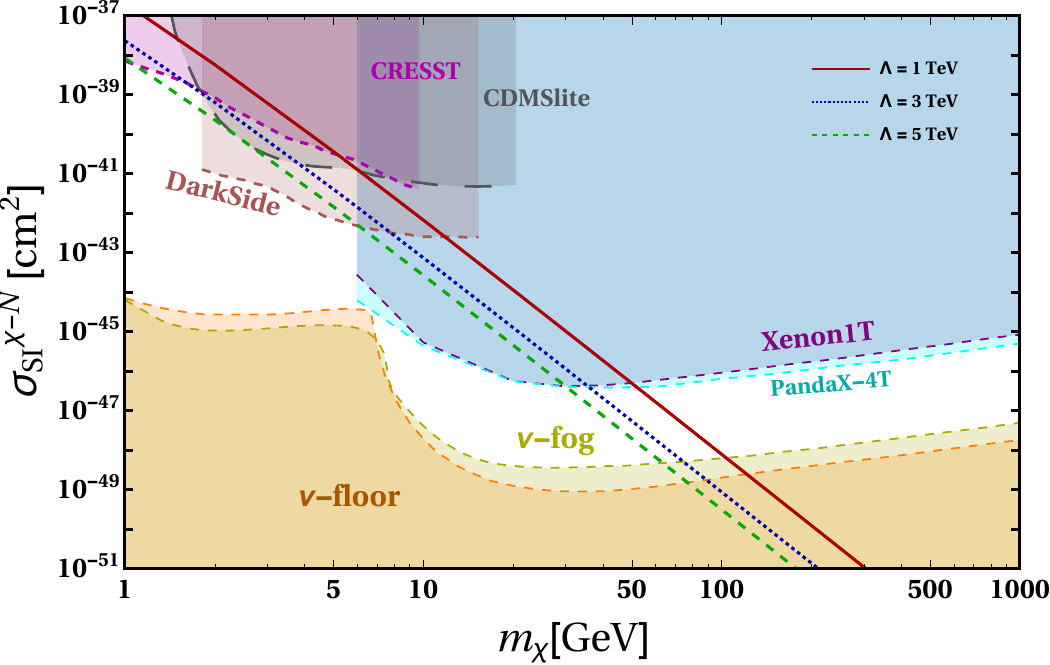}
    \caption{The DM-nucleon spin-independent elastic scattering cross-section as a function of the dark matter mass ($m_\chi$) is shown for different values of the cutoff scale $\Lambda$. The shaded regions are ruled out by different direct-detection experiments.}
    \label{fig:DD_CS_Plot2}
\end{figure}

\section{Constraints from Direct Detection Experiments}
The most stringent constraint on the DM parameter space comes from the direct detection experiments. Therefore, it is important to discuss the outlook of the DDM in direct detection experiments. In \autoref{fig:DD_CS_Plot2} we have shown the DM-nucleon spin-independent elastic scattering cross-section ($\sigma_{SI}^{\chi-N}$) as a function of the DM mass ($m_\chi$) for different values of the cutoff scale, $\Lambda$. The cross-sections are estimated using the \texttt{micrOMEGAs}~\cite{Belanger:2008sj, Belanger:2020gnr} package assuming equal form factors of the protons and neutrons. In the plot, we only show the lines for the $\mathcal{C}_M=\mathcal{C}_E=1$ case as we noted that the interaction with only the electric dipole moment (i.e., $\mathcal{C}_M=0$) vanishes resulting in unphysical cross-section. We presented the current exclusion limits for DM-nucleon spin-independent elastic scattering from some of the leading direct-detection experiments for the DM mass range of $[1, 1000]\,\mathrm{GeV}$ with the shaded regions. The purple and cyan regions for respectively, Xenon1T~\cite{XENON:2018voc}, and PandaX-4T~\cite{PandaX-4T:2021bab} for DM mass 6 GeV to 1000 GeV are the strongest ranging from $10^{-44}\,\mathrm{cm}^2$ to $10^{-48}\,\mathrm{cm}^2$. Note that the experimental limits from the DM-nucleon interactions dominate over the DM-electron interactions. Therefore, we have not included the limits for DM-electron scattering in the plot. For the lower mass region, we included the constraints from DarkSide \cite{DarkSide:2018bpj}, CRESST \cite{CRESST:2015txj}, CDMSlite \cite{SuperCDMS:2017nns} for which the cross-section ranges from $10^{-38}\,\mathrm{cm}^2$ to $10^{-42}\,\mathrm{cm}^2$. The orange-shaded region at the bottom of the plot arises due to the similarity of nuclear recoil signature between DM of different masses and neutrino from different sources~\cite{Billard:2013cxa} and is termed as \emph{neutrino floor}. However, the strength of this neutrino background comes from the systematic uncertainties on their fluxes and therefore it is not impermeable. In fact, efforts are being made to penetrate this region of the DM parameter space~\cite{Ruppin:2014bra, Davis:2014ama, Sassi:2021umf}. Terming this region as \emph{neutrino fog}, ref.~\cite{OHare:2021utq} introduced an updated definition, which is shown by the yellow-shaded region in the plot. So, we find that while the ideal mass-window of DM observation is $\sim30\,$GeV to $\sim 100\,$GeV depending on the value of $\Lambda$ if the neutrino background can be distinguished we might be able to probe WIMPs with mass up to $\sim300\,$GeV. 
\begin{figure}[t]
\centering
\includegraphics[scale=0.76]{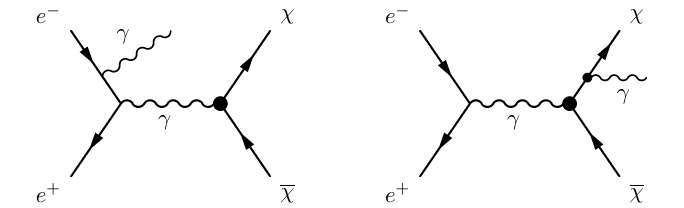}\\
(a)\hspace{4.4cm}(b)
\caption{Feynman diagram for the mono-photon signal at an $e^+e^-$ collider. The photon can be radiated off either the electron or the positron leg.}
\label{fig:mono-g_Feyn}
\end{figure}

\section{Collider Phenomenology} \label{sec:3}

Assuming that the dark matter particles $\chi$ can be produced in the $e^+e^-$ collision process, it can be observed through the mono-photon signature. \autoref{fig:mono-g_Feyn} shows the representative Feynman diagrams for such processes $(e^+e^- \rightarrow \chi \Bar{\chi} \gamma )$ where the photon can come from the initial state leptons or the final state DM particle legs. Since the dark matter particle $\chi$ escapes the detector contributing to missing energy, we need a visible particle to identify the events. This purpose is fulfilled by the initial state radiation (ISR) and final state radiation (FSR) photon. 

\begin{figure}[t]
\centering
\includegraphics[width=0.9\linewidth]{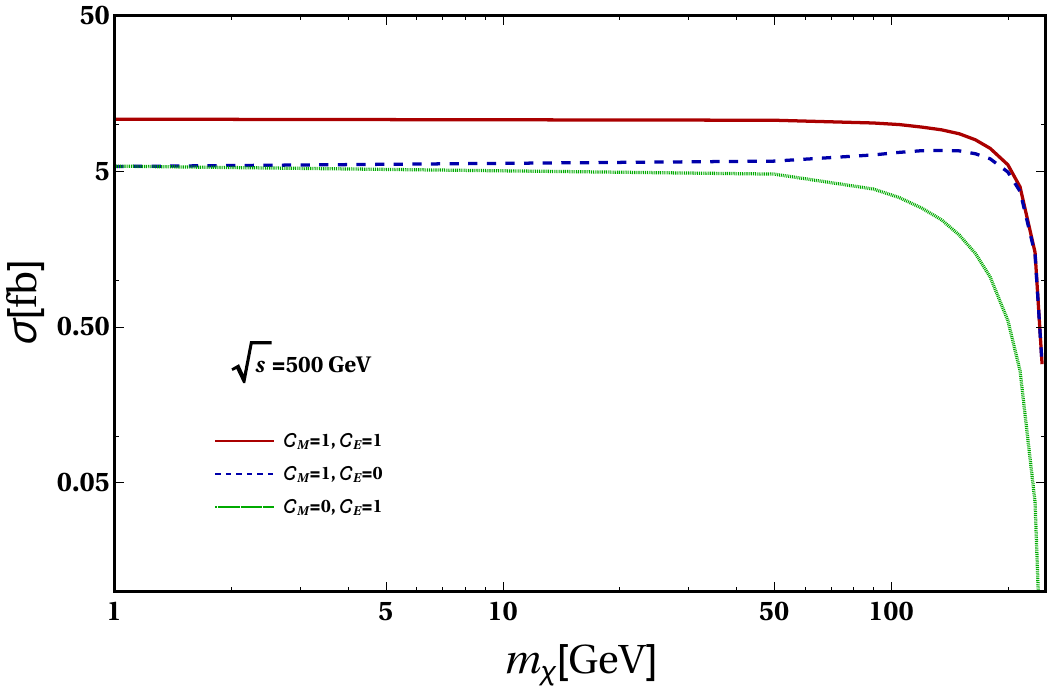}
\caption{Variation of mono-photon signal cross section (fb) with the DM mass $m_{\chi}$ (GeV) for different cases of $(\mathcal{C}_M, \mathcal{C}_E)$ as $(1,1),(1,0),(0,1)$ by red, green and blue lines respectively for $\sqrt{s} = 500$ GeV.}
\label{fig:CSplot}
\end{figure}

In \autoref{fig:CSplot} we have shown the variation of the cross-section with respect to the dark matter mass($m_\chi$) for all three cases of $(\mathcal{C}_M, \mathcal{C}_E)$ corresponding to the machine c.o.m energy $\sqrt{s} = 500~\rm{GeV}$. Two background processes that potentially can contribute to such mono-photon searches --- the radiative neutrino pair production $e^+e^-\to\nu\overline{\nu}\gamma$ and radiative Bhabha scattering $e^+e^-\to e^+e^-\gamma$. The neutrino pair background is the dominant irreducible and exactly mimics the mono-photon signals. It is highly polarization-dependent and thus can be suppressed significantly by polarizing the incoming beams. On the other hand, the Bhabha scattering contributes to the mono-photon background only when the final state electron-positron pair escapes detection. It is not affected by beam polarization but can be reduced by improving the detection capability of $e^+e^-$-pairs in the forward direction. In the next section, we describe at length the details of event simulation and analysis for both unpolarized and polarized beam cases.

\subsection{Event Simulation}
\begin{table*}[ht!]
\caption{\label{table:Pol}%
Comparison of the irreducible SM backgrounds and signal cross sections (merged) for different choices of beam polarization at $\sqrt{s} = 500~\rm{GeV}$. For the signal, we have chosen benchmark values of $m_\chi = 50~\rm{GeV}$ and $\Lambda = 3~\rm{TeV}$. The numbers in bold highlight the optimal polarization choice for a given operator type.}
\begin{ruledtabular}
\begin{tabular}{lcccccc} 
\textbf{Process} & \textbf{Unpolarized} & \textbf{Polarization} &  \multicolumn{4}{c}{\textbf{Polarized cross-section (fb)}}   \\
  \cline{4-7}
 \textbf{type} & \textbf{CS (fb)} & \textbf{\boldmath{$P(e^{-},e^{+})$}} & $(+,+)$ & $(+,-)$ & $(-,+)$ & $(-,-)$ \\
\colrule
           $\nu\overline{\nu}+N\gamma$                     &5581.0& $(\phantom{.}80,0\phantom{.})$ & 1629.8 & 1629.8 & 9571.4 & 9571.4 \\
        &  & $(80,30)$ & 1770.9 & 1478.3 & 12412.9 & 6737.3 \\ \hline
                      $e^+e^-+N\gamma$          &255608& $(\phantom{.}80,0\phantom{.})$ &255156  &255156  &258463  &258463  \\
         &  & $(80,30)$ & 254348 & 254299 & 256423 & 252165 \\ \hline
                  $C_M=C_E=1$          &10.67& $(\phantom{.}80,0\phantom{.})$ & 10.67 & 10.67 & 10.63 & 10.63 \\
       &  & $(80,30)$ & 8.09 & \textbf{13.22} & 13.16 & 8.09 \\ \hline
                  $C_M=1, \; C_E=0$     &5.81& $(\phantom{.}80,0\phantom{.})$ & 5.81 & 5.81 & 5.82 & 5.82 \\
         &  & $(80,30)$ & 4.41 & \textbf{7.22} & 7.20 & 4.41 \\ \hline
                    $C_M=0, \; C_E=1$   &4.82& $(\phantom{.}80,0\phantom{.})$ & 4.82 & 4.82 & 4.82 & 4.82 \\
        &  & $(80,30)$ & 3.66 & \textbf{5.97} & 5.98 & 3.67 \\
\end{tabular}
\end{ruledtabular}
\end{table*}

We implemented our dipole DM model in \texttt{FeynRules}~\cite{Alloul:2013bka} to obtain the UFO library files which were then used in \texttt{Whizard} \cite{Kilian:2007gr} to estimate the cross-sections of the signal. The background processes are also generated using Whizard with both the beamstrahlung and ISR effects. These are two important beam-induced backgrounds that muddle the signal photons and dilute the signals. Hence, careful modeling of these beamstrahlung and ISR photons is imperative for a mono-photon analysis at $e^-e^+$ colliders. We implement beamstrahlung using a beam spectra file generated using \texttt{circe2} for ILC. For ISR modeling, we used the `recoil' mode of the handler with the ISR mass equal to the electron mass. \texttt{Whizard} gives us freedom in selecting the generation level cuts that help in removing the infrared and collinear singularities. To remove these divergences in the cross-section calculations we restricted the event generation phase space with the following generation-level cuts : 
\begin{equation}
p_T^{\gamma}>2\text{ GeV},\;\;\; E_{\gamma}>1\text{ GeV},\;\;\; |\cos{\theta_{\gamma}|} \leq 0.9975
\label{eq:gencut}   
\end{equation}

To avoid soft and collinear divergences in radiative Bhabha scattering, a few more restrictions were applied:
\begin{align}
    &M_{e_{in}^{\pm},e_{out}^{\pm}} < -2\text{ GeV} ,\;\; M_{e_{out}^{\pm},e_{out}^{\pm}} > 2\text{ GeV},\nonumber\\ 
    &M_{e^{\pm},\gamma_i} > 2\text{ GeV} ,\;\; M_{e^{\pm},\gamma} > 4\text{ GeV}  
\end{align}
 where $\gamma_i$ and $\gamma$ denote matrix element (ME) photon and signal photon, respectively. In order to take into account the proper summation of higher-order corrections and avoid double counting, we have used the ME-ISR merging technique introduced in \cite{Kalinowski:2020lhp}. We considered two independent variables $q_+$ and $q_-$ as introduced in \cite{Kalinowski:2020lhp} and are defined as :
 \begin{eqnarray}
      q_- &=\sqrt{2E_{\rm cm}E_{\gamma}}\sin\frac{\theta_{\gamma}}{2} \nonumber \\
      q_+ &=\sqrt{2E_{\rm cm}E_{\gamma}}\cos\frac{\theta_{\gamma}}{2}
     \label{eq:merge-q}
 \end{eqnarray}
where $E_{cm}$ = $\sqrt{s}$, $E_{\gamma}$ and $\theta_{\gamma}$ are the center of mass energy, photon energy, and photon polar angle respectively. 
The rejection cuts used in our analysis for ME photons are  $ q_\pm < 1~\rm{GeV}$ or $E_\gamma < 1 ~\rm{GeV}$ and for ISR photons, they are $ q_\pm > 1~\rm{GeV}$ and $E_\gamma > 1~\rm{GeV}$, respectively. 

We have used \texttt{Delphes3} \cite{deFavereau:2013fsa} with \texttt{ILCgen} configuration card for the fast detector simulations of the generated signal and background events with the above-discussed cuts. 
It can be seen from \autoref{fig:CSplot} that there is almost no change in the unpolarized signal cross-section up to a certain value of dark matter mass and after that, it decreases rapidly due to phase space suppression for all three cases of $(\mathcal{C}_M,\mathcal{C}_E)$ as $(1,1),(1,0),(0,1)$. 
 
\subsection{Effect of polarization}
As per the ILC baseline design, we have studied the effect of polarization on the incoming $e^-$ and $e^+$ beams on the signal and background cross-sections by considering two cases of polarization configurations. First, we examined the effect of polarization on only the electron beam. Then we studied how the application of polarization on the positron beam also affects the signal and background cross-sections. In \autoref{table:Pol}, we presented the cross-sections of the two backgrounds and the signals with all four possible helicity orientations, namely, $(+,+),(+,-),(-,+)$ and $(-,-)$ in each case \cite{Barklow:2015tja, Kundu:2021cmo}. From \autoref{table:Pol} we can see that the neutrino background cross-section is highly polarization dependent, which can best be reduced to $26.49\%$ by choosing $(+80,-30)$ as the polarization choice of incoming beams, whereas the Bhabha background shows a polarization independent behaviour. Also, the signal can be moderately enhanced by using the same polarization choice for all cases of $(\mathcal{C}_M, \mathcal{C}_E)$ values as $(1,1),\;(1,0),\;(0,1)$ by $23.90\%$, $24.27\%$ and $23.86\%$ respectively with an integrated luminosity of 4 ab$^{-1}$. Hence, we will consider $(+80,-30)$ as our polarization choice for further analysis in this paper.We have also considered the H20 scenario for comparison which considers shared polarization states of $40\%$ for $(+,-)$ and $(-,+)$  and $10\%$ for $(+,+)$ and $(-,-)$  configuration at $\sqrt{s}=500GeV$.

\begin{figure*}[ht]
    \centering
    \includegraphics[width=0.4\linewidth]{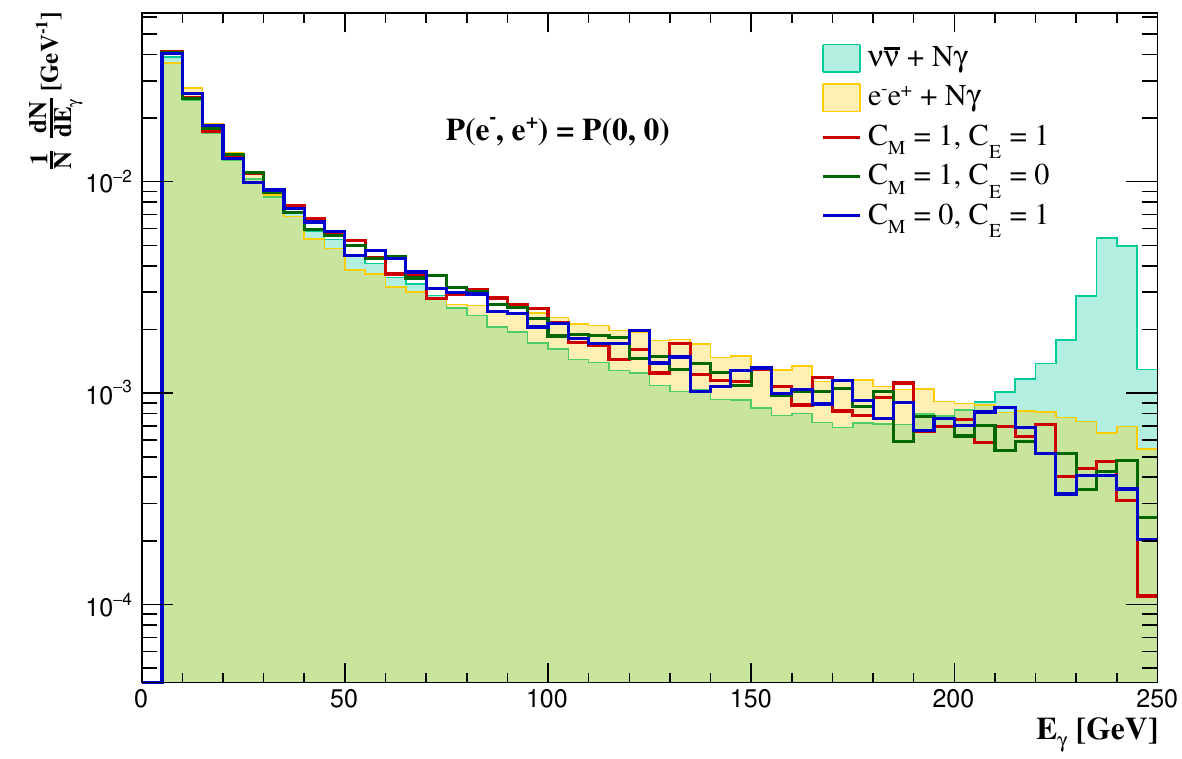}\hspace{25pt}
    \includegraphics[width=0.4\linewidth]{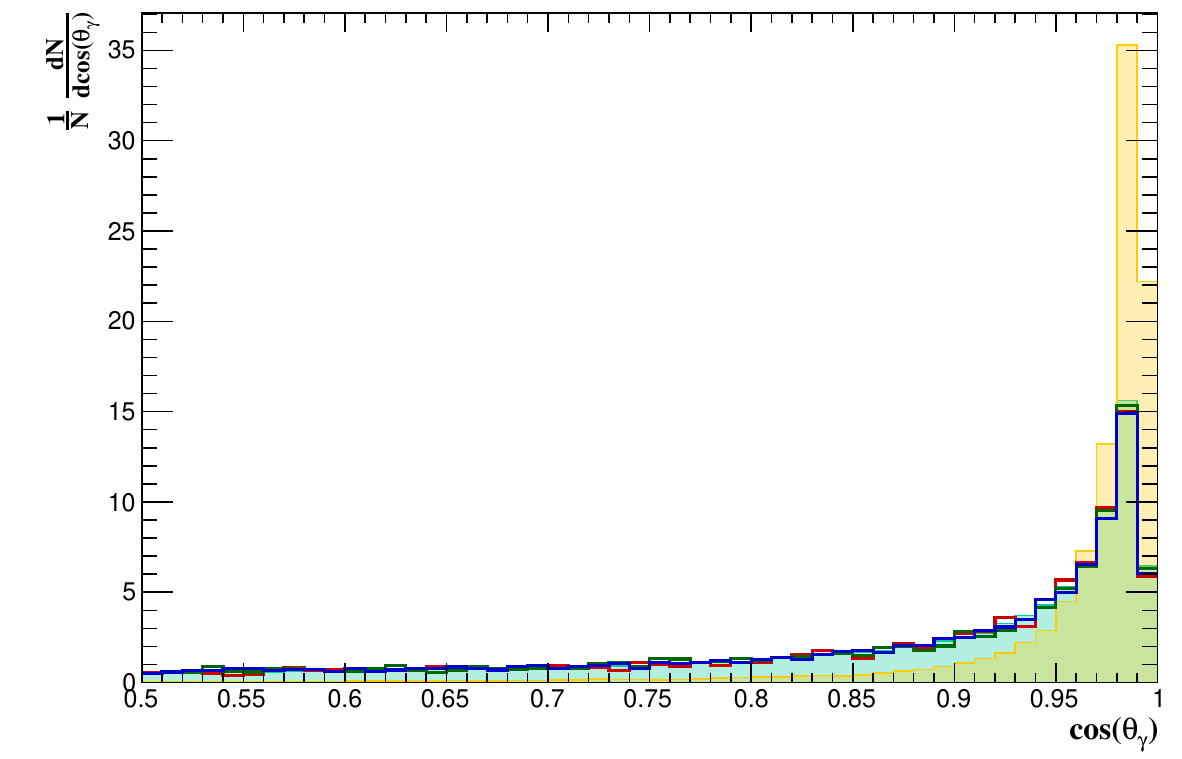}
    \caption{Normalized differential distributions after baseline selection cuts for $E_\gamma$, $\cos\theta_{\gamma}$ for the unpolarized beam case for DM mass of $m_\chi=1$ GeV and $\Lambda = 3$ TeV. In each panel, the shaded blue histograms are for the neutrino background and the shaded yellow histograms are for the radiative Bhabha background. The unshaded red, green, and blue histograms are for the three scenarios of $(\mathcal{C}_M, \mathcal{C}_E)$ as $(1,1),\;(1,0),\;(0,1)$, respectively.}
    \label{fig:histos}
\end{figure*}

\subsection{Event Selection}
In this section, we discuss the cut-based analysis performed in the \texttt{ROOT}~\cite{BRUN199781} framework with the detector-simulated signal and background datasets to get an optimized value of the signal-to-background ratio. For that purpose we have considered here benchmark values of $m_{\chi}=1$ GeV and $\Lambda=1$ TeV although, in the next section, we will vary the DM mass to get $3\sigma$ sensitivity limit on the cutoff scale $\Lambda$.\par

In the below, we described different selection cuts we have used in order to minimize the contribution of the two backgrounds.

\noindent\textbf{Baseline selection cuts ---} Keeping in mind the configuration of the generic International Large Detector (ILD) and Silicon Detector (SiD) at ILC we set the baseline selection criteria. We define our mono-photon signal events that contain at least one photon that satisfies the following criteria (\autoref{table:cuts} ) 
\begin{equation}
    E_{\gamma}>5\text{ GeV} \; , \; |\eta_{\gamma}| < 2.8
    \label{eq:baseline}
\end{equation}
where, the $E_\gamma$ and $\eta_\gamma$ are the photon energy and pseudorapidity. We assume here that the tracking detector has a range of $7^\circ-173^\circ$ in the polar angle. In the case of multiple photons in an event, the one with the highest transverse momentum is considered the signal photon. In \autoref{fig:histos} we have shown the normalized distributions of the photon energy and pseudorapidity after the baseline-selection cut. 

On top of the baseline -selection criteria we used some additional cuts 

\noindent\textbf{Cut 1 ---} We can reduce neutrino background by putting an appropriate constraint on the value of $E_{\gamma}$ which can be worked out by analyzing the left panel of \autoref{fig:histos}. We can see that the neutrino background peaks at around $E_{\gamma}=235$ GeV because of the radiative return of Z-resonance, whereas the signal almost vanishes at $E_{\gamma}=212$ GeV. Thus, we can reduce both backgrounds by requiring $E_{\gamma}<210$ GeV. \par

\noindent\textbf{Cut 2 ---} The baseline cut on pseudorapidity $|\eta_{\gamma}| < 2.8 $ is to account for the inability of the detector to detect collinear photons outside the range of $7^\circ < \theta < 173^\circ$. But in order to decrease the background signal, we have applied a better cut on the quantity $|\cos\theta_{\gamma}|<0.95$ which can be seen in the right panel of the \autoref{fig:histos}. The Bhabha background peaks to a much higher value than the signal in the region  $|\cos\theta_{\gamma}|> 0.95$ and should be excluded. \par

\noindent\textbf{Cut 3 ---} We restrict $ p_T > 2 $ GeV for the charged particles that are reconstructed as electrons. This criterion was also used in Ref.\cite{Habermehl:2018yul} and reduces the Bhabha event yield.
\par

\noindent\textbf{Cut 4 ---} The BeamCal veto is used to reject those events that have deposited energy in the forward direction of the beamline in an electromagnetic calorimeter which detects highly energetic electrons that escape the central tracker and hence suppress the contribution from the Bhabha background events \cite{Abramowicz:2010bg}.

\begin{table}[t]
\caption{\label{table:cuts} Event selection criteria applied to the signal and Standard Model(SM) backgrounds. }
\begin{ruledtabular}
\begin{tabular}{lc} 
  Description & Definition \\ 
  \colrule
  Baseline selection & $E_{\gamma} > 5\text{ GeV},\;\; |\eta_{\gamma}| < 2.8,$ \\
  Cut-1 & $E_{\gamma} < 210~\rm{GeV}$ \\
  Cut-2 & $|\cos\theta_{\gamma}| < 0.95$ \\
  Cut-3 & \phantom{00}Charged particle veto with $p_{T} > 5 $~GeV \\
  Cut-4 &  BeamCal veto \\
\end{tabular}
\end{ruledtabular}
\end{table}
\subsection{Signal significance} 
There can be several sources of systematic uncertainties at the ILC like, beam polarization, event estimation, integrated luminosity, and the shape of luminosity spectrum \cite{Habermehl:2020njb,Kalinowski:2021tyr,Habermehl:2018yul} which should be considered for realistic uncertainty analysis along with there correlations which are beyond the scope of our work and hence, we have done a simplified analysis by considering only background estimation uncertainty in a cut-and-count method by considering uncertainties of $0.2 \%$ and $1 \%$ for neutrino-pair and Bhabha background respectively \cite{Kalinowski:2020lhp,Habermehl:2018yul,Blaising:2021vhh}. We expect that the experiments will show sensitivities that will lie somewhere in between the two scenarios we have considered, i.e., with no systematics and with our choice of systematics as stated above.

\begin{figure*}[htb]
    \centering
    \includegraphics[width=0.46\linewidth]{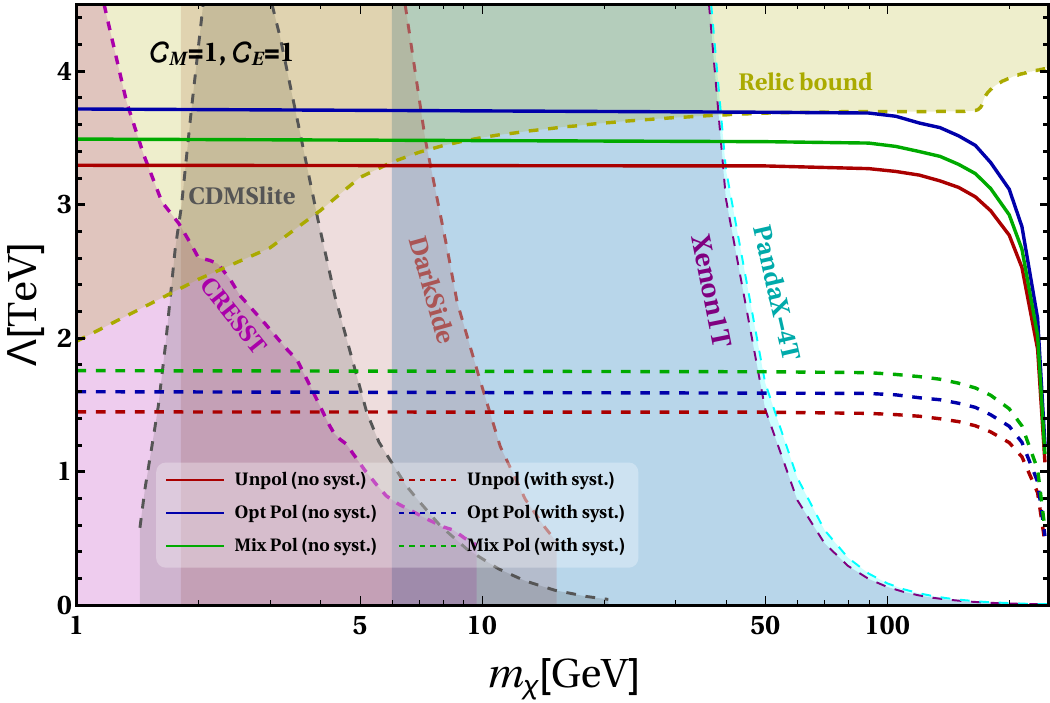}\hspace{10pt}
    \includegraphics[width=0.46\linewidth]{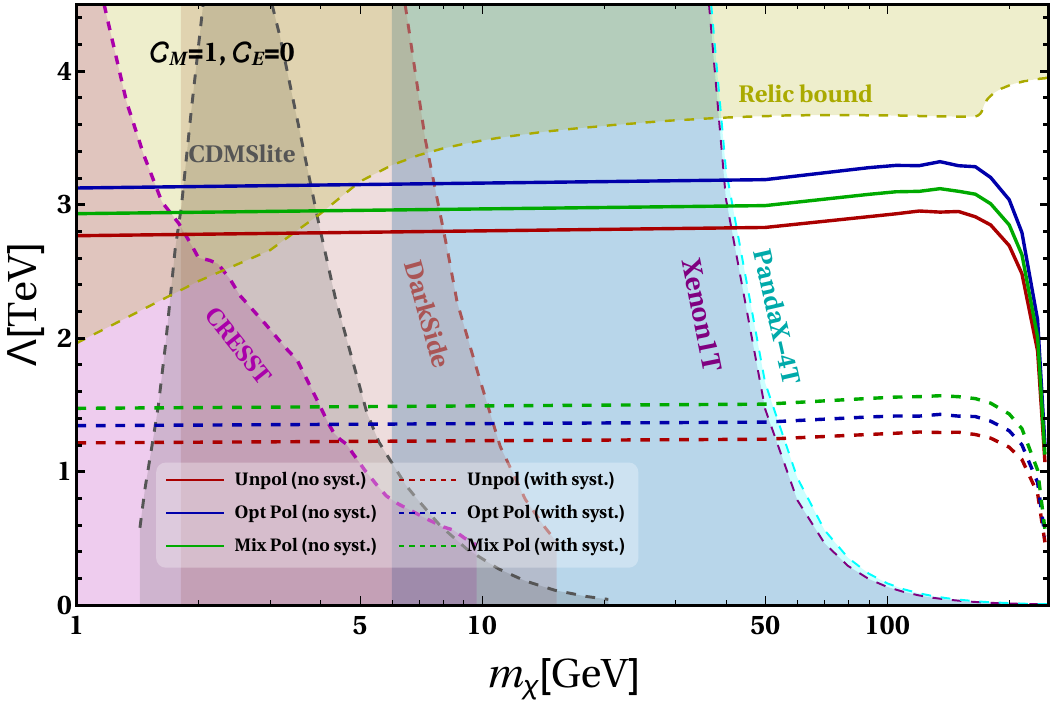}\vspace{15pt}
    \includegraphics[width=0.46\linewidth]{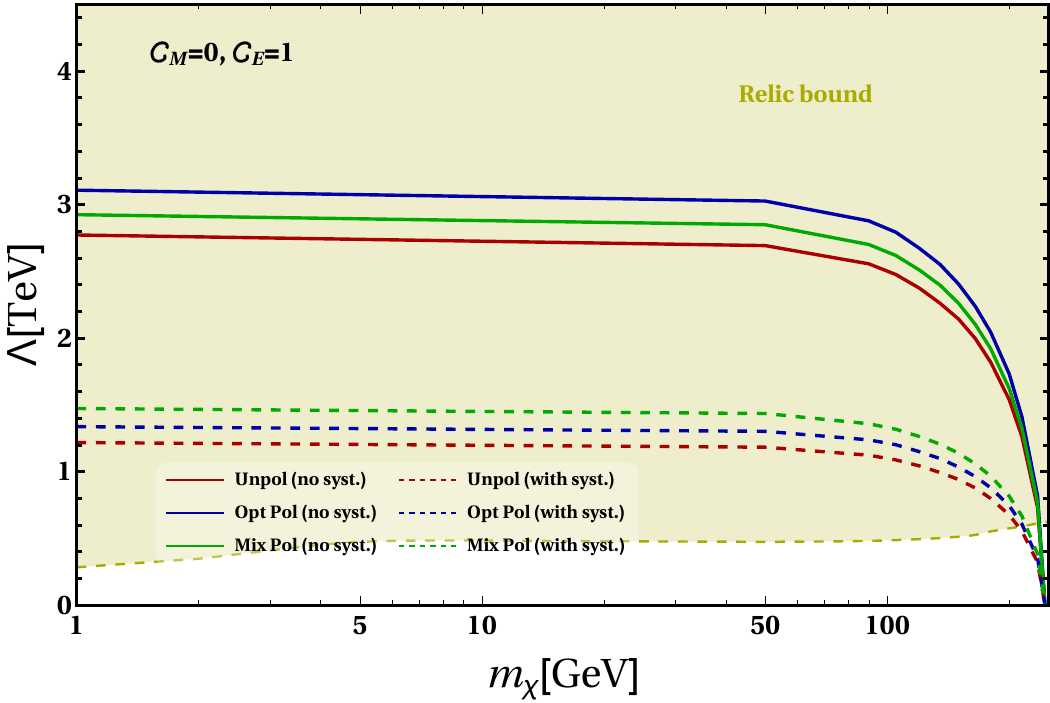}
    \caption{$3\sigma$ sensitivity contours in the mono-photon channel for the three combinations of the coupling values with \emph{unpolarized} (red lines), \emph{optimally polarized} (blue lines) and \emph{mixed polarized} (red lines) $e^+e^-$ beams at $\sqrt s=500$~GeV center-of-mass energy and with  ${\cal L}_{\rm int}=4$ ab$^{-1}$ integrated luminosity. The solid (dashed) contours are without (with) background systematics.}
    \label{fig:contours}
\end{figure*}

The signal significance is calculated using 
\begin{equation}
    Sig=\frac{S}{\sqrt{S+\Sigma_i(B_i+\epsilon_i^2B_i^2)}}
\end{equation}
where S denotes the number of signal events $B_i$ denotes $i$-th channel background events for a given value of integrated luminosity, and the corresponding systematic uncertainty in background estimation is given by $\epsilon_i$.For the mixed polarization (H20) scenario  which  considers shared polarization states of $40\%$ for $(+,-)$ and $(-,+)$  and $10\%$ for $(+,+)$ and $(-,-)$  configuration at $\sqrt{s}=500$ GeV, significance for individual polarization case has been calculated (assuming them to be uncorrelated) using

\begin{equation}
    Sig_{H20} = \sqrt{{Sig^2_{(+,+)}}+{Sig^2_{(+,-)}}+{Sig^2_{(-,+)}}+{Sig^2_{(-,-)}}}.
\end{equation}

\subsection{Results and Summary}\label{sec:4}
The $3\sigma$ sensitivity contours in the $m_\chi-\Lambda$ plane in the mono-photon channel for the three combinations of the coupling values $(\mathcal{C}_M,\mathcal{C}_E)$ as $(1,1),(1,0),(0,1)$ with \emph{unpolarized} (red lines), \emph{optimally polarized} (blue lines), and \emph{mixed polarized} (green lines) at 500 GeV ILC with 4 ab$^{-1}$ of integrated luminosity without (with) systematics are shown in \autoref{fig:contours} with solid (dashed) lines. In the plots, we have also shown the direct detection bound on $\Lambda$ as a function of $m_\chi$ that follows from CRESST, CDMSlite, DarkSide, Xenon1T, and PandaX-4T amongst which the most stringent bounds are given by Xenon1T and PandaX-4T with the shaded regions. The dark yellow dashed lines represent the values of the parameters that gives the correct Planck-observed relic density and the shaded region above results in overabundance of DM and therefore is forbidden. Hence, the unshaded region under the contour lines of our DDM model shows the allowed region where the DDM with the correct relic density can be observed.

\begin{table}[htb]
\caption{\label{tab:my_label} $3\sigma$ upper limit on the cutoff scale at 500 GeV ILC with 4 ab$^{-1}$ of integrated luminosity without (with) systematics for three different cases --- with both EDM and MDM are non-zero, with only MDM, and with only EDM. Three beam polarization scenarios are considered namely, unpolarized, optimally polarized (Opt. Pol.), and mixed polarization case of H20 scenario (Mix. Pol.).}
\begin{ruledtabular}
\begin{tabular}{lccc} 
    & \multicolumn{3}{l}{Upper limit on $\Lambda$ (TeV) for $m_\chi=1$ GeV at}  \\ 
    Type & \multicolumn{3}{l}{$\mathcal{L}_{int}=4\text{ ab}^{-1}$} \\\cline{2-4}
    & Unpolarized & Opt. Pol. & Mix. Pol. \\ 
    \colrule
     $\mathcal{C}_M=1,~~\mathcal{C}_E=1$ & 3.30 (1.45) & 3.72 (1.60) & 3.49 (1.76)\\
     $\mathcal{C}_M=1,~~\mathcal{C}_E=0$ & 2.95 (1.30) & 3.32 (1.43) & 3.12 (1.57) \\
     $\mathcal{C}_M=0,~~\mathcal{C}_E=1$ & 2.77 (1.22) & 3.12 (1.34) & 2.92 (1.47) \\
    \end{tabular}
\end{ruledtabular}
\end{table}
In  \autoref{tab:my_label}, we have displayed the $3\sigma$ upper limit on the cutoff scale $\Lambda$ corresponding to $m_\chi = 1$ GeV at 500 GeV ILC with 4 ab$^{-1}$ of integrated luminosity without (with) systematics for three different cases mentioned above along with three beam polarization scenarios namely, unpolarized, optimally polarized (Opt. Pol.) and mixed polarized case of H20 scenario (Mix. Pol.). We found that in all three coupling combinations of $(\mathcal{C}_M,\mathcal{C}_E)$ as $(1,1), (1,0),(0,1)$, the best bounds are given by the Optimally polarized case when we do not consider the systematics and those are $3.14, 2.59$ and $2.59$ TeV respectively. With systematics, the H20 scenario is found to give better bounds of $1.48,1.22$ and $1.23 $ TeV respectively in comparison to Unpolarized and Optimally polarized cases.

Finally, we observed that among the three scenarios of incoming beam polarization, namely, the unpolarized beam, optimally polarized beam, and maximally polarized beam, the best constraint on the cut-off scale $\Lambda$ arises from the optimally polarized beam configuration. Specifically, for $(\mathcal{C}_M, \mathcal{C}_E)$ equal to $(1,1)$ and without considering systematics, we have determined this constraint to be $3.72$ TeV. However, when taking systematics into consideration, we obtain a relatively relaxed bound of $1.76$ TeV for $(\mathcal{C}_M, \mathcal{C}_E)$  set to $(1,1)$ in the case of mixed polarization.

\bibliographystyle{apsrev4-1}
\bibliography{Reference.bib}

\begin{thebibliography}{64}%
\makeatletter
\providecommand \@ifxundefined [1]{%
 \@ifx{#1\undefined}
}%
\providecommand \@ifnum [1]{%
 \ifnum #1\expandafter \@firstoftwo
 \else \expandafter \@secondoftwo
 \fi
}%
\providecommand \@ifx [1]{%
 \ifx #1\expandafter \@firstoftwo
 \else \expandafter \@secondoftwo
 \fi
}%
\providecommand \natexlab [1]{#1}%
\providecommand \enquote  [1]{``#1''}%
\providecommand \bibnamefont  [1]{#1}%
\providecommand \bibfnamefont [1]{#1}%
\providecommand \citenamefont [1]{#1}%
\providecommand \href@noop [0]{\@secondoftwo}%
\providecommand \href [0]{\begingroup \@sanitize@url \@href}%
\providecommand \@href[1]{\@@startlink{#1}\@@href}%
\providecommand \@@href[1]{\endgroup#1\@@endlink}%
\providecommand \@sanitize@url [0]{\catcode `\\12\catcode `\$12\catcode
  `\&12\catcode `\#12\catcode `\^12\catcode `\_12\catcode `\%12\relax}%
\providecommand \@@startlink[1]{}%
\providecommand \@@endlink[0]{}%
\providecommand \url  [0]{\begingroup\@sanitize@url \@url }%
\providecommand \@url [1]{\endgroup\@href {#1}{\urlprefix }}%
\providecommand \urlprefix  [0]{URL }%
\providecommand \Eprint [0]{\href }%
\providecommand \doibase [0]{http://dx.doi.org/}%
\providecommand \selectlanguage [0]{\@gobble}%
\providecommand \bibinfo  [0]{\@secondoftwo}%
\providecommand \bibfield  [0]{\@secondoftwo}%
\providecommand \translation [1]{[#1]}%
\providecommand \BibitemOpen [0]{}%
\providecommand \bibitemStop [0]{}%
\providecommand \bibitemNoStop [0]{.\EOS\space}%
\providecommand \EOS [0]{\spacefactor3000\relax}%
\providecommand \BibitemShut  [1]{\csname bibitem#1\endcsname}%
\let\auto@bib@innerbib\@empty
\bibitem [{\citenamefont {Sofue}\ and\ \citenamefont
  {Rubin}(2001)}]{Sofue:2000jx}%
  \BibitemOpen
  \bibfield  {author} {\bibinfo {author} {\bibfnamefont {Y.}~\bibnamefont
  {Sofue}}\ and\ \bibinfo {author} {\bibfnamefont {V.}~\bibnamefont {Rubin}},\
  }\href {\doibase 10.1146/annurev.astro.39.1.137} {\bibfield  {journal}
  {\bibinfo  {journal} {Ann. Rev. Astron. Astrophys.}\ }\textbf {\bibinfo
  {volume} {39}},\ \bibinfo {pages} {137} (\bibinfo {year} {2001})},\ \Eprint
  {http://arxiv.org/abs/astro-ph/0010594} {arXiv:astro-ph/0010594} \BibitemShut
  {NoStop}%
\bibitem [{\citenamefont {Chang}\ and\ \citenamefont
  {Necib}(2021)}]{Chang:2020rem}%
  \BibitemOpen
  \bibfield  {author} {\bibinfo {author} {\bibfnamefont {L.~J.}\ \bibnamefont
  {Chang}}\ and\ \bibinfo {author} {\bibfnamefont {L.}~\bibnamefont {Necib}},\
  }\href {\doibase 10.1093/mnras/stab2440} {\bibfield  {journal} {\bibinfo
  {journal} {Mon. Not. Roy. Astron. Soc.}\ }\textbf {\bibinfo {volume} {507}},\
  \bibinfo {pages} {4715} (\bibinfo {year} {2021})},\ \Eprint
  {http://arxiv.org/abs/2009.00613} {arXiv:2009.00613 [astro-ph.CO]}
  \BibitemShut {NoStop}%
\bibitem [{\citenamefont {Zwicky}(1937)}]{Zwicky:1937zza}%
  \BibitemOpen
  \bibfield  {author} {\bibinfo {author} {\bibfnamefont {F.}~\bibnamefont
  {Zwicky}},\ }\href {\doibase 10.1086/143864} {\bibfield  {journal} {\bibinfo
  {journal} {Astrophys. J.}\ }\textbf {\bibinfo {volume} {86}},\ \bibinfo
  {pages} {217} (\bibinfo {year} {1937})}\BibitemShut {NoStop}%
\bibitem [{\citenamefont {Clowe}\ \emph {et~al.}(2006)\citenamefont {Clowe},
  \citenamefont {Bradac}, \citenamefont {Gonzalez}, \citenamefont {Markevitch},
  \citenamefont {Randall}, \citenamefont {Jones},\ and\ \citenamefont
  {Zaritsky}}]{Clowe:2006eq}%
  \BibitemOpen
  \bibfield  {author} {\bibinfo {author} {\bibfnamefont {D.}~\bibnamefont
  {Clowe}}, \bibinfo {author} {\bibfnamefont {M.}~\bibnamefont {Bradac}},
  \bibinfo {author} {\bibfnamefont {A.~H.}\ \bibnamefont {Gonzalez}}, \bibinfo
  {author} {\bibfnamefont {M.}~\bibnamefont {Markevitch}}, \bibinfo {author}
  {\bibfnamefont {S.~W.}\ \bibnamefont {Randall}}, \bibinfo {author}
  {\bibfnamefont {C.}~\bibnamefont {Jones}}, \ and\ \bibinfo {author}
  {\bibfnamefont {D.}~\bibnamefont {Zaritsky}},\ }\href {\doibase
  10.1086/508162} {\bibfield  {journal} {\bibinfo  {journal} {Astrophys. J.
  Lett.}\ }\textbf {\bibinfo {volume} {648}},\ \bibinfo {pages} {L109}
  (\bibinfo {year} {2006})},\ \Eprint {http://arxiv.org/abs/astro-ph/0608407}
  {arXiv:astro-ph/0608407} \BibitemShut {NoStop}%
\bibitem [{\citenamefont {Massey}\ \emph {et~al.}(2010)\citenamefont {Massey},
  \citenamefont {Kitching},\ and\ \citenamefont {Richard}}]{Massey:2010hh}%
  \BibitemOpen
  \bibfield  {author} {\bibinfo {author} {\bibfnamefont {R.}~\bibnamefont
  {Massey}}, \bibinfo {author} {\bibfnamefont {T.}~\bibnamefont {Kitching}}, \
  and\ \bibinfo {author} {\bibfnamefont {J.}~\bibnamefont {Richard}},\ }\href
  {\doibase 10.1088/0034-4885/73/8/086901} {\bibfield  {journal} {\bibinfo
  {journal} {Rept. Prog. Phys.}\ }\textbf {\bibinfo {volume} {73}},\ \bibinfo
  {pages} {086901} (\bibinfo {year} {2010})},\ \Eprint
  {http://arxiv.org/abs/1001.1739} {arXiv:1001.1739 [astro-ph.CO]} \BibitemShut
  {NoStop}%
\bibitem [{\citenamefont {Bennett}\ \emph {et~al.}(2013)\citenamefont
  {Bennett}, \citenamefont {Larson}, \citenamefont {Weiland}, \citenamefont
  {Jarosik}, \citenamefont {Hinshaw}, \citenamefont {Odegard}, \citenamefont
  {Smith}, \citenamefont {Hill}, \citenamefont {Gold}, \citenamefont {Halpern},
  \citenamefont {Komatsu}, \citenamefont {Nolta}, \citenamefont {Page},
  \citenamefont {Spergel}, \citenamefont {Wollack}, \citenamefont {Dunkley},
  \citenamefont {Kogut}, \citenamefont {Limon}, \citenamefont {Meyer},
  \citenamefont {Tucker},\ and\ \citenamefont {Wright}}]{Bennett_2013}%
  \BibitemOpen
  \bibfield  {author} {\bibinfo {author} {\bibfnamefont {C.~L.}\ \bibnamefont
  {Bennett}}, \bibinfo {author} {\bibfnamefont {D.}~\bibnamefont {Larson}},
  \bibinfo {author} {\bibfnamefont {J.~L.}\ \bibnamefont {Weiland}}, \bibinfo
  {author} {\bibfnamefont {N.}~\bibnamefont {Jarosik}}, \bibinfo {author}
  {\bibfnamefont {G.}~\bibnamefont {Hinshaw}}, \bibinfo {author} {\bibfnamefont
  {N.}~\bibnamefont {Odegard}}, \bibinfo {author} {\bibfnamefont {K.~M.}\
  \bibnamefont {Smith}}, \bibinfo {author} {\bibfnamefont {R.~S.}\ \bibnamefont
  {Hill}}, \bibinfo {author} {\bibfnamefont {B.}~\bibnamefont {Gold}}, \bibinfo
  {author} {\bibfnamefont {M.}~\bibnamefont {Halpern}}, \bibinfo {author}
  {\bibfnamefont {E.}~\bibnamefont {Komatsu}}, \bibinfo {author} {\bibfnamefont
  {M.~R.}\ \bibnamefont {Nolta}}, \bibinfo {author} {\bibfnamefont
  {L.}~\bibnamefont {Page}}, \bibinfo {author} {\bibfnamefont {D.~N.}\
  \bibnamefont {Spergel}}, \bibinfo {author} {\bibfnamefont {E.}~\bibnamefont
  {Wollack}}, \bibinfo {author} {\bibfnamefont {J.}~\bibnamefont {Dunkley}},
  \bibinfo {author} {\bibfnamefont {A.}~\bibnamefont {Kogut}}, \bibinfo
  {author} {\bibfnamefont {M.}~\bibnamefont {Limon}}, \bibinfo {author}
  {\bibfnamefont {S.~S.}\ \bibnamefont {Meyer}}, \bibinfo {author}
  {\bibfnamefont {G.~S.}\ \bibnamefont {Tucker}}, \ and\ \bibinfo {author}
  {\bibfnamefont {E.~L.}\ \bibnamefont {Wright}},\ }\href {\doibase
  10.1088/0067-0049/208/2/20} {\bibfield  {journal} {\bibinfo  {journal} {The
  Astrophysical Journal Supplement Series}\ }\textbf {\bibinfo {volume}
  {208}},\ \bibinfo {pages} {20} (\bibinfo {year} {2013})}\BibitemShut
  {NoStop}%
\bibitem [{\citenamefont {Aghanim}\ \emph {et~al.}(2020)\citenamefont {Aghanim}
  \emph {et~al.}}]{Planck:2018vyg}%
  \BibitemOpen
  \bibfield  {author} {\bibinfo {author} {\bibfnamefont {N.}~\bibnamefont
  {Aghanim}} \emph {et~al.} (\bibinfo {collaboration} {Planck}),\ }\href
  {\doibase 10.1051/0004-6361/201833910} {\bibfield  {journal} {\bibinfo
  {journal} {Astron. Astrophys.}\ }\textbf {\bibinfo {volume} {641}},\ \bibinfo
  {pages} {A6} (\bibinfo {year} {2020})},\ \bibinfo {note} {[Erratum:
  Astron.Astrophys. 652, C4 (2021)]},\ \Eprint
  {http://arxiv.org/abs/1807.06209} {arXiv:1807.06209 [astro-ph.CO]}
  \BibitemShut {NoStop}%
\bibitem [{\citenamefont {Billard}\ \emph
  {et~al.}(2014{\natexlab{a}})\citenamefont {Billard}, \citenamefont
  {Strigari},\ and\ \citenamefont {Figueroa-Feliciano}}]{Billard:2013qya}%
  \BibitemOpen
  \bibfield  {author} {\bibinfo {author} {\bibfnamefont {J.}~\bibnamefont
  {Billard}}, \bibinfo {author} {\bibfnamefont {L.}~\bibnamefont {Strigari}}, \
  and\ \bibinfo {author} {\bibfnamefont {E.}~\bibnamefont
  {Figueroa-Feliciano}},\ }\href {\doibase 10.1103/PhysRevD.89.023524}
  {\bibfield  {journal} {\bibinfo  {journal} {Phys. Rev. D}\ }\textbf {\bibinfo
  {volume} {89}},\ \bibinfo {pages} {023524} (\bibinfo {year}
  {2014}{\natexlab{a}})},\ \Eprint {http://arxiv.org/abs/1307.5458}
  {arXiv:1307.5458 [hep-ph]} \BibitemShut {NoStop}%
\bibitem [{\citenamefont
  {O'Hare}(2021{\natexlab{a}})}]{PhysRevLett.127.251802}%
  \BibitemOpen
  \bibfield  {author} {\bibinfo {author} {\bibfnamefont {C.~A.~J.}\
  \bibnamefont {O'Hare}},\ }\href {\doibase 10.1103/PhysRevLett.127.251802}
  {\bibfield  {journal} {\bibinfo  {journal} {Phys. Rev. Lett.}\ }\textbf
  {\bibinfo {volume} {127}},\ \bibinfo {pages} {251802} (\bibinfo {year}
  {2021}{\natexlab{a}})}\BibitemShut {NoStop}%
\bibitem [{\citenamefont {Bernal}\ \emph {et~al.}(2017)\citenamefont {Bernal},
  \citenamefont {Heikinheimo}, \citenamefont {Tenkanen}, \citenamefont
  {Tuominen},\ and\ \citenamefont {Vaskonen}}]{Bernal:2017kxu}%
  \BibitemOpen
  \bibfield  {author} {\bibinfo {author} {\bibfnamefont {N.}~\bibnamefont
  {Bernal}}, \bibinfo {author} {\bibfnamefont {M.}~\bibnamefont {Heikinheimo}},
  \bibinfo {author} {\bibfnamefont {T.}~\bibnamefont {Tenkanen}}, \bibinfo
  {author} {\bibfnamefont {K.}~\bibnamefont {Tuominen}}, \ and\ \bibinfo
  {author} {\bibfnamefont {V.}~\bibnamefont {Vaskonen}},\ }\href {\doibase
  10.1142/S0217751X1730023X} {\bibfield  {journal} {\bibinfo  {journal} {Int.
  J. Mod. Phys. A}\ }\textbf {\bibinfo {volume} {32}},\ \bibinfo {pages}
  {1730023} (\bibinfo {year} {2017})},\ \Eprint
  {http://arxiv.org/abs/1706.07442} {arXiv:1706.07442 [hep-ph]} \BibitemShut
  {NoStop}%
\bibitem [{\citenamefont {Emken}\ \emph {et~al.}(2019)\citenamefont {Emken},
  \citenamefont {Essig}, \citenamefont {Kouvaris},\ and\ \citenamefont
  {Sholapurkar}}]{Emken:2019tni}%
  \BibitemOpen
  \bibfield  {author} {\bibinfo {author} {\bibfnamefont {T.}~\bibnamefont
  {Emken}}, \bibinfo {author} {\bibfnamefont {R.}~\bibnamefont {Essig}},
  \bibinfo {author} {\bibfnamefont {C.}~\bibnamefont {Kouvaris}}, \ and\
  \bibinfo {author} {\bibfnamefont {M.}~\bibnamefont {Sholapurkar}},\ }\href
  {\doibase 10.1088/1475-7516/2019/09/070} {\bibfield  {journal} {\bibinfo
  {journal} {JCAP}\ }\textbf {\bibinfo {volume} {09}},\ \bibinfo {pages} {070}
  (\bibinfo {year} {2019})},\ \Eprint {http://arxiv.org/abs/1905.06348}
  {arXiv:1905.06348 [hep-ph]} \BibitemShut {NoStop}%
\bibitem [{\citenamefont {Mavromatos}\ \emph {et~al.}(2017)\citenamefont
  {Mavromatos}, \citenamefont {Arg\"uelles}, \citenamefont {Ruffini},\ and\
  \citenamefont {Rueda}}]{Mavromatos:2017lfs}%
  \BibitemOpen
  \bibfield  {author} {\bibinfo {author} {\bibfnamefont {N.~E.}\ \bibnamefont
  {Mavromatos}}, \bibinfo {author} {\bibfnamefont {C.~R.}\ \bibnamefont
  {Arg\"uelles}}, \bibinfo {author} {\bibfnamefont {R.}~\bibnamefont
  {Ruffini}}, \ and\ \bibinfo {author} {\bibfnamefont {J.~A.}\ \bibnamefont
  {Rueda}},\ }in\ \href {\doibase 10.1142/9789813226609_0035} {\emph {\bibinfo
  {booktitle} {{14th Marcel Grossmann Meeting on Recent Developments in
  Theoretical and Experimental General Relativity, Astrophysics, and
  Relativistic Field Theories}}}},\ Vol.~\bibinfo {volume} {1}\ (\bibinfo
  {year} {2017})\ pp.\ \bibinfo {pages} {639--666}\BibitemShut {NoStop}%
\bibitem [{\citenamefont {Stebbins}\ and\ \citenamefont
  {Krnjaic}(2019)}]{Stebbins:2019xjr}%
  \BibitemOpen
  \bibfield  {author} {\bibinfo {author} {\bibfnamefont {A.}~\bibnamefont
  {Stebbins}}\ and\ \bibinfo {author} {\bibfnamefont {G.}~\bibnamefont
  {Krnjaic}},\ }\href {\doibase 10.1088/1475-7516/2019/12/003} {\bibfield
  {journal} {\bibinfo  {journal} {JCAP}\ }\textbf {\bibinfo {volume} {12}},\
  \bibinfo {pages} {003} (\bibinfo {year} {2019})},\ \Eprint
  {http://arxiv.org/abs/1908.05275} {arXiv:1908.05275 [astro-ph.CO]}
  \BibitemShut {NoStop}%
\bibitem [{\citenamefont {Davidson}\ \emph {et~al.}(2000)\citenamefont
  {Davidson}, \citenamefont {Hannestad},\ and\ \citenamefont
  {Raffelt}}]{Davidson:2000hf}%
  \BibitemOpen
  \bibfield  {author} {\bibinfo {author} {\bibfnamefont {S.}~\bibnamefont
  {Davidson}}, \bibinfo {author} {\bibfnamefont {S.}~\bibnamefont {Hannestad}},
  \ and\ \bibinfo {author} {\bibfnamefont {G.}~\bibnamefont {Raffelt}},\ }\href
  {\doibase 10.1088/1126-6708/2000/05/003} {\bibfield  {journal} {\bibinfo
  {journal} {JHEP}\ }\textbf {\bibinfo {volume} {05}},\ \bibinfo {pages} {003}
  (\bibinfo {year} {2000})},\ \Eprint {http://arxiv.org/abs/hep-ph/0001179}
  {arXiv:hep-ph/0001179} \BibitemShut {NoStop}%
\bibitem [{\citenamefont {Hambye}\ and\ \citenamefont
  {Xu}(2021)}]{Hambye:2021xvd}%
  \BibitemOpen
  \bibfield  {author} {\bibinfo {author} {\bibfnamefont {T.}~\bibnamefont
  {Hambye}}\ and\ \bibinfo {author} {\bibfnamefont {X.-J.}\ \bibnamefont
  {Xu}},\ }\href {\doibase 10.1007/JHEP11(2021)156} {\bibfield  {journal}
  {\bibinfo  {journal} {JHEP}\ }\textbf {\bibinfo {volume} {11}},\ \bibinfo
  {pages} {156} (\bibinfo {year} {2021})},\ \Eprint
  {http://arxiv.org/abs/2106.01403} {arXiv:2106.01403 [hep-ph]} \BibitemShut
  {NoStop}%
\bibitem [{\citenamefont {Sigurdson}\ \emph {et~al.}(2004)\citenamefont
  {Sigurdson}, \citenamefont {Doran}, \citenamefont {Kurylov}, \citenamefont
  {Caldwell},\ and\ \citenamefont {Kamionkowski}}]{Sigurdson:2004zp}%
  \BibitemOpen
  \bibfield  {author} {\bibinfo {author} {\bibfnamefont {K.}~\bibnamefont
  {Sigurdson}}, \bibinfo {author} {\bibfnamefont {M.}~\bibnamefont {Doran}},
  \bibinfo {author} {\bibfnamefont {A.}~\bibnamefont {Kurylov}}, \bibinfo
  {author} {\bibfnamefont {R.~R.}\ \bibnamefont {Caldwell}}, \ and\ \bibinfo
  {author} {\bibfnamefont {M.}~\bibnamefont {Kamionkowski}},\ }\href {\doibase
  10.1103/PhysRevD.70.083501} {\bibfield  {journal} {\bibinfo  {journal} {Phys.
  Rev. D}\ }\textbf {\bibinfo {volume} {70}},\ \bibinfo {pages} {083501}
  (\bibinfo {year} {2004})},\ \bibinfo {note} {[Erratum: Phys.Rev.D 73, 089903
  (2006)]},\ \Eprint {http://arxiv.org/abs/astro-ph/0406355}
  {arXiv:astro-ph/0406355} \BibitemShut {NoStop}%
\bibitem [{\citenamefont {Mahmoudi}\ \emph {et~al.}(2020)\citenamefont
  {Mahmoudi}, \citenamefont {Haghighat}, \citenamefont {Vamegh},\ and\
  \citenamefont {Mohammadi}}]{Mahmoudi:2018lll}%
  \BibitemOpen
  \bibfield  {author} {\bibinfo {author} {\bibfnamefont {S.}~\bibnamefont
  {Mahmoudi}}, \bibinfo {author} {\bibfnamefont {M.}~\bibnamefont {Haghighat}},
  \bibinfo {author} {\bibfnamefont {S.~A.~M.}\ \bibnamefont {Vamegh}}, \ and\
  \bibinfo {author} {\bibfnamefont {R.}~\bibnamefont {Mohammadi}},\ }\href
  {\doibase 10.1140/epjc/s10052-020-7982-y} {\bibfield  {journal} {\bibinfo
  {journal} {Eur. Phys. J. C}\ }\textbf {\bibinfo {volume} {80}},\ \bibinfo
  {pages} {402} (\bibinfo {year} {2020})},\ \Eprint
  {http://arxiv.org/abs/1805.11172} {arXiv:1805.11172 [hep-ph]} \BibitemShut
  {NoStop}%
\bibitem [{\citenamefont {Barger}\ \emph
  {et~al.}(2011{\natexlab{a}})\citenamefont {Barger}, \citenamefont {Keung},\
  and\ \citenamefont {Marfatia}}]{BARGER201174}%
  \BibitemOpen
  \bibfield  {author} {\bibinfo {author} {\bibfnamefont {V.}~\bibnamefont
  {Barger}}, \bibinfo {author} {\bibfnamefont {W.-Y.}\ \bibnamefont {Keung}}, \
  and\ \bibinfo {author} {\bibfnamefont {D.}~\bibnamefont {Marfatia}},\ }\href
  {\doibase https://doi.org/10.1016/j.physletb.2010.12.008} {\bibfield
  {journal} {\bibinfo  {journal} {Physics Letters B}\ }\textbf {\bibinfo
  {volume} {696}},\ \bibinfo {pages} {74} (\bibinfo {year}
  {2011}{\natexlab{a}})}\BibitemShut {NoStop}%
\bibitem [{\citenamefont {Geytenbeek}\ \emph {et~al.}(2017)\citenamefont
  {Geytenbeek}, \citenamefont {Rao}, \citenamefont {Scott}, \citenamefont
  {Serenelli}, \citenamefont {Vincent}, \citenamefont {White},\ and\
  \citenamefont {Williams}}]{Geytenbeek:2016nfg}%
  \BibitemOpen
  \bibfield  {author} {\bibinfo {author} {\bibfnamefont {B.}~\bibnamefont
  {Geytenbeek}}, \bibinfo {author} {\bibfnamefont {S.}~\bibnamefont {Rao}},
  \bibinfo {author} {\bibfnamefont {P.}~\bibnamefont {Scott}}, \bibinfo
  {author} {\bibfnamefont {A.}~\bibnamefont {Serenelli}}, \bibinfo {author}
  {\bibfnamefont {A.~C.}\ \bibnamefont {Vincent}}, \bibinfo {author}
  {\bibfnamefont {M.}~\bibnamefont {White}}, \ and\ \bibinfo {author}
  {\bibfnamefont {A.~G.}\ \bibnamefont {Williams}},\ }\href {\doibase
  10.1088/1475-7516/2017/03/029} {\bibfield  {journal} {\bibinfo  {journal}
  {JCAP}\ }\textbf {\bibinfo {volume} {03}},\ \bibinfo {pages} {029} (\bibinfo
  {year} {2017})},\ \Eprint {http://arxiv.org/abs/1610.06737} {arXiv:1610.06737
  [hep-ph]} \BibitemShut {NoStop}%
\bibitem [{\citenamefont {Binh}\ \emph {et~al.}(2020)\citenamefont {Binh},
  \citenamefont {Binh},\ and\ \citenamefont {Long}}]{Binh:2020xtf}%
  \BibitemOpen
  \bibfield  {author} {\bibinfo {author} {\bibfnamefont {D.~T.}\ \bibnamefont
  {Binh}}, \bibinfo {author} {\bibfnamefont {V.~H.}\ \bibnamefont {Binh}}, \
  and\ \bibinfo {author} {\bibfnamefont {H.~N.}\ \bibnamefont {Long}},\
  }\href@noop {} {\  (\bibinfo {year} {2020})},\ \Eprint
  {http://arxiv.org/abs/2006.09020} {arXiv:2006.09020 [hep-ph]} \BibitemShut
  {NoStop}%
\bibitem [{\citenamefont {Aranda}\ \emph {et~al.}(2016)\citenamefont {Aranda},
  \citenamefont {Barajas},\ and\ \citenamefont {Cembranos}}]{Aranda:2015jis}%
  \BibitemOpen
  \bibfield  {author} {\bibinfo {author} {\bibfnamefont {A.}~\bibnamefont
  {Aranda}}, \bibinfo {author} {\bibfnamefont {L.}~\bibnamefont {Barajas}}, \
  and\ \bibinfo {author} {\bibfnamefont {J.~A.~R.}\ \bibnamefont {Cembranos}},\
  }\href {\doibase 10.1088/1475-7516/2016/03/034} {\bibfield  {journal}
  {\bibinfo  {journal} {JCAP}\ }\textbf {\bibinfo {volume} {03}},\ \bibinfo
  {pages} {034} (\bibinfo {year} {2016})},\ \Eprint
  {http://arxiv.org/abs/1511.02805} {arXiv:1511.02805 [hep-ph]} \BibitemShut
  {NoStop}%
\bibitem [{\citenamefont {Lopes}\ \emph {et~al.}(2014)\citenamefont {Lopes},
  \citenamefont {Kadota},\ and\ \citenamefont {Silk}}]{Lopes:2013xua}%
  \BibitemOpen
  \bibfield  {author} {\bibinfo {author} {\bibfnamefont {I.}~\bibnamefont
  {Lopes}}, \bibinfo {author} {\bibfnamefont {K.}~\bibnamefont {Kadota}}, \
  and\ \bibinfo {author} {\bibfnamefont {J.}~\bibnamefont {Silk}},\ }\href
  {\doibase 10.1088/2041-8205/780/2/L15} {\bibfield  {journal} {\bibinfo
  {journal} {Astrophys. J. Lett.}\ }\textbf {\bibinfo {volume} {780}},\
  \bibinfo {pages} {L15} (\bibinfo {year} {2014})},\ \Eprint
  {http://arxiv.org/abs/1310.0673} {arXiv:1310.0673 [astro-ph.SR]} \BibitemShut
  {NoStop}%
\bibitem [{\citenamefont {Mass\'o}\ \emph {et~al.}(2009)\citenamefont
  {Mass\'o}, \citenamefont {Mohanty},\ and\ \citenamefont
  {Rao}}]{PhysRevD.80.036009}%
  \BibitemOpen
  \bibfield  {author} {\bibinfo {author} {\bibfnamefont {E.}~\bibnamefont
  {Mass\'o}}, \bibinfo {author} {\bibfnamefont {S.}~\bibnamefont {Mohanty}}, \
  and\ \bibinfo {author} {\bibfnamefont {S.}~\bibnamefont {Rao}},\ }\href
  {\doibase 10.1103/PhysRevD.80.036009} {\bibfield  {journal} {\bibinfo
  {journal} {Phys. Rev. D}\ }\textbf {\bibinfo {volume} {80}},\ \bibinfo
  {pages} {036009} (\bibinfo {year} {2009})}\BibitemShut {NoStop}%
\bibitem [{\citenamefont {Chang}\ \emph {et~al.}(2021)\citenamefont {Chang},
  \citenamefont {Essig},\ and\ \citenamefont {Reinert}}]{Chang:2019xva}%
  \BibitemOpen
  \bibfield  {author} {\bibinfo {author} {\bibfnamefont {J.~H.}\ \bibnamefont
  {Chang}}, \bibinfo {author} {\bibfnamefont {R.}~\bibnamefont {Essig}}, \ and\
  \bibinfo {author} {\bibfnamefont {A.}~\bibnamefont {Reinert}},\ }\href
  {\doibase 10.1007/JHEP03(2021)141} {\bibfield  {journal} {\bibinfo  {journal}
  {JHEP}\ }\textbf {\bibinfo {volume} {03}},\ \bibinfo {pages} {141} (\bibinfo
  {year} {2021})},\ \Eprint {http://arxiv.org/abs/1911.03389} {arXiv:1911.03389
  [hep-ph]} \BibitemShut {NoStop}%
\bibitem [{\citenamefont {Barger}\ \emph
  {et~al.}(2011{\natexlab{b}})\citenamefont {Barger}, \citenamefont {Keung},\
  and\ \citenamefont {Marfatia}}]{Barger:2010gv}%
  \BibitemOpen
  \bibfield  {author} {\bibinfo {author} {\bibfnamefont {V.}~\bibnamefont
  {Barger}}, \bibinfo {author} {\bibfnamefont {W.-Y.}\ \bibnamefont {Keung}}, \
  and\ \bibinfo {author} {\bibfnamefont {D.}~\bibnamefont {Marfatia}},\ }\href
  {\doibase 10.1016/j.physletb.2010.12.008} {\bibfield  {journal} {\bibinfo
  {journal} {Phys. Lett. B}\ }\textbf {\bibinfo {volume} {696}},\ \bibinfo
  {pages} {74} (\bibinfo {year} {2011}{\natexlab{b}})},\ \Eprint
  {http://arxiv.org/abs/1007.4345} {arXiv:1007.4345 [hep-ph]} \BibitemShut
  {NoStop}%
\bibitem [{\citenamefont {Fitzpatrick}\ and\ \citenamefont
  {Zurek}(2010)}]{Fitzpatrick:2010br}%
  \BibitemOpen
  \bibfield  {author} {\bibinfo {author} {\bibfnamefont {A.~L.}\ \bibnamefont
  {Fitzpatrick}}\ and\ \bibinfo {author} {\bibfnamefont {K.~M.}\ \bibnamefont
  {Zurek}},\ }\href {\doibase 10.1103/PhysRevD.82.075004} {\bibfield  {journal}
  {\bibinfo  {journal} {Phys. Rev. D}\ }\textbf {\bibinfo {volume} {82}},\
  \bibinfo {pages} {075004} (\bibinfo {year} {2010})},\ \Eprint
  {http://arxiv.org/abs/1007.5325} {arXiv:1007.5325 [hep-ph]} \BibitemShut
  {NoStop}%
\bibitem [{\citenamefont {Banks}\ \emph {et~al.}(2010)\citenamefont {Banks},
  \citenamefont {Fortin},\ and\ \citenamefont {Thomas}}]{Banks:2010eh}%
  \BibitemOpen
  \bibfield  {author} {\bibinfo {author} {\bibfnamefont {T.}~\bibnamefont
  {Banks}}, \bibinfo {author} {\bibfnamefont {J.-F.}\ \bibnamefont {Fortin}}, \
  and\ \bibinfo {author} {\bibfnamefont {S.}~\bibnamefont {Thomas}},\
  }\href@noop {} {\  (\bibinfo {year} {2010})},\ \Eprint
  {http://arxiv.org/abs/1007.5515} {arXiv:1007.5515 [hep-ph]} \BibitemShut
  {NoStop}%
\bibitem [{\citenamefont {Del~Nobile}\ \emph {et~al.}(2012)\citenamefont
  {Del~Nobile}, \citenamefont {Kouvaris}, \citenamefont {Panci}, \citenamefont
  {Sannino},\ and\ \citenamefont {Virkajarvi}}]{DelNobile:2012tx}%
  \BibitemOpen
  \bibfield  {author} {\bibinfo {author} {\bibfnamefont {E.}~\bibnamefont
  {Del~Nobile}}, \bibinfo {author} {\bibfnamefont {C.}~\bibnamefont
  {Kouvaris}}, \bibinfo {author} {\bibfnamefont {P.}~\bibnamefont {Panci}},
  \bibinfo {author} {\bibfnamefont {F.}~\bibnamefont {Sannino}}, \ and\
  \bibinfo {author} {\bibfnamefont {J.}~\bibnamefont {Virkajarvi}},\ }\href
  {\doibase 10.1088/1475-7516/2012/08/010} {\bibfield  {journal} {\bibinfo
  {journal} {JCAP}\ }\textbf {\bibinfo {volume} {08}},\ \bibinfo {pages} {010}
  (\bibinfo {year} {2012})},\ \Eprint {http://arxiv.org/abs/1203.6652}
  {arXiv:1203.6652 [hep-ph]} \BibitemShut {NoStop}%
\bibitem [{\citenamefont {Kopp}\ \emph {et~al.}(2014)\citenamefont {Kopp},
  \citenamefont {Michaels},\ and\ \citenamefont {Smirnov}}]{Kopp:2014tsa}%
  \BibitemOpen
  \bibfield  {author} {\bibinfo {author} {\bibfnamefont {J.}~\bibnamefont
  {Kopp}}, \bibinfo {author} {\bibfnamefont {L.}~\bibnamefont {Michaels}}, \
  and\ \bibinfo {author} {\bibfnamefont {J.}~\bibnamefont {Smirnov}},\ }\href
  {\doibase 10.1088/1475-7516/2014/04/022} {\bibfield  {journal} {\bibinfo
  {journal} {JCAP}\ }\textbf {\bibinfo {volume} {04}},\ \bibinfo {pages} {022}
  (\bibinfo {year} {2014})},\ \Eprint {http://arxiv.org/abs/1401.6457}
  {arXiv:1401.6457 [hep-ph]} \BibitemShut {NoStop}%
\bibitem [{\citenamefont {Chatterjee}\ and\ \citenamefont
  {Laha}(2023)}]{Chatterjee:2022gbo}%
  \BibitemOpen
  \bibfield  {author} {\bibinfo {author} {\bibfnamefont {S.}~\bibnamefont
  {Chatterjee}}\ and\ \bibinfo {author} {\bibfnamefont {R.}~\bibnamefont
  {Laha}},\ }\href {\doibase 10.1103/PhysRevD.107.083036} {\bibfield  {journal}
  {\bibinfo  {journal} {Phys. Rev. D}\ }\textbf {\bibinfo {volume} {107}},\
  \bibinfo {pages} {083036} (\bibinfo {year} {2023})},\ \Eprint
  {http://arxiv.org/abs/2202.13339} {arXiv:2202.13339 [hep-ph]} \BibitemShut
  {NoStop}%
\bibitem [{\citenamefont {Flacke}\ and\ \citenamefont
  {Maybury}(2007)}]{Flacke:2006ut}%
  \BibitemOpen
  \bibfield  {author} {\bibinfo {author} {\bibfnamefont {T.}~\bibnamefont
  {Flacke}}\ and\ \bibinfo {author} {\bibfnamefont {D.~W.}\ \bibnamefont
  {Maybury}},\ }\href {\doibase 10.1142/S0218271807010985} {\bibfield
  {journal} {\bibinfo  {journal} {Int. J. Mod. Phys. D}\ }\textbf {\bibinfo
  {volume} {16}},\ \bibinfo {pages} {1593} (\bibinfo {year} {2007})},\ \Eprint
  {http://arxiv.org/abs/hep-ph/0601161} {arXiv:hep-ph/0601161} \BibitemShut
  {NoStop}%
\bibitem [{\citenamefont {Kadota}\ and\ \citenamefont
  {Silk}(2014)}]{Kadota:2014mea}%
  \BibitemOpen
  \bibfield  {author} {\bibinfo {author} {\bibfnamefont {K.}~\bibnamefont
  {Kadota}}\ and\ \bibinfo {author} {\bibfnamefont {J.}~\bibnamefont {Silk}},\
  }\href {\doibase 10.1103/PhysRevD.89.103528} {\bibfield  {journal} {\bibinfo
  {journal} {Phys. Rev. D}\ }\textbf {\bibinfo {volume} {89}},\ \bibinfo
  {pages} {103528} (\bibinfo {year} {2014})},\ \Eprint
  {http://arxiv.org/abs/1402.7295} {arXiv:1402.7295 [hep-ph]} \BibitemShut
  {NoStop}%
\bibitem [{\citenamefont {Dienes}\ \emph {et~al.}(2023)\citenamefont {Dienes},
  \citenamefont {Feng}, \citenamefont {Fieg}, \citenamefont {Huang},
  \citenamefont {Lee},\ and\ \citenamefont {Thomas}}]{Dienes:2023uve}%
  \BibitemOpen
  \bibfield  {author} {\bibinfo {author} {\bibfnamefont {K.~R.}\ \bibnamefont
  {Dienes}}, \bibinfo {author} {\bibfnamefont {J.~L.}\ \bibnamefont {Feng}},
  \bibinfo {author} {\bibfnamefont {M.}~\bibnamefont {Fieg}}, \bibinfo {author}
  {\bibfnamefont {F.}~\bibnamefont {Huang}}, \bibinfo {author} {\bibfnamefont
  {S.~J.}\ \bibnamefont {Lee}}, \ and\ \bibinfo {author} {\bibfnamefont
  {B.}~\bibnamefont {Thomas}},\ }\href {\doibase 10.1103/PhysRevD.107.115006}
  {\bibfield  {journal} {\bibinfo  {journal} {Phys. Rev. D}\ }\textbf {\bibinfo
  {volume} {107}},\ \bibinfo {pages} {115006} (\bibinfo {year} {2023})},\
  \Eprint {http://arxiv.org/abs/2301.05252} {arXiv:2301.05252 [hep-ph]}
  \BibitemShut {NoStop}%
\bibitem [{\citenamefont {Kling}\ \emph {et~al.}(2023)\citenamefont {Kling},
  \citenamefont {Kuo}, \citenamefont {Trojanowski},\ and\ \citenamefont
  {Tsai}}]{Kling:2022ykt}%
  \BibitemOpen
  \bibfield  {author} {\bibinfo {author} {\bibfnamefont {F.}~\bibnamefont
  {Kling}}, \bibinfo {author} {\bibfnamefont {J.-L.}\ \bibnamefont {Kuo}},
  \bibinfo {author} {\bibfnamefont {S.}~\bibnamefont {Trojanowski}}, \ and\
  \bibinfo {author} {\bibfnamefont {Y.-D.}\ \bibnamefont {Tsai}},\ }\href
  {\doibase 10.1016/j.nuclphysb.2023.116103} {\bibfield  {journal} {\bibinfo
  {journal} {Nucl. Phys. B}\ }\textbf {\bibinfo {volume} {987}},\ \bibinfo
  {pages} {116103} (\bibinfo {year} {2023})},\ \Eprint
  {http://arxiv.org/abs/2205.09137} {arXiv:2205.09137 [hep-ph]} \BibitemShut
  {NoStop}%
\bibitem [{\citenamefont {Fortin}\ and\ \citenamefont
  {Tait}(2012)}]{Fortin:2011hv}%
  \BibitemOpen
  \bibfield  {author} {\bibinfo {author} {\bibfnamefont {J.-F.}\ \bibnamefont
  {Fortin}}\ and\ \bibinfo {author} {\bibfnamefont {T.~M.~P.}\ \bibnamefont
  {Tait}},\ }\href {\doibase 10.1103/PhysRevD.85.063506} {\bibfield  {journal}
  {\bibinfo  {journal} {Phys. Rev. D}\ }\textbf {\bibinfo {volume} {85}},\
  \bibinfo {pages} {063506} (\bibinfo {year} {2012})},\ \Eprint
  {http://arxiv.org/abs/1103.3289} {arXiv:1103.3289 [hep-ph]} \BibitemShut
  {NoStop}%
\bibitem [{\citenamefont {Kahlhoefer}(2017)}]{Kahlhoefer:2017dnp}%
  \BibitemOpen
  \bibfield  {author} {\bibinfo {author} {\bibfnamefont {F.}~\bibnamefont
  {Kahlhoefer}},\ }\href {\doibase 10.1142/S0217751X1730006X} {\bibfield
  {journal} {\bibinfo  {journal} {Int. J. Mod. Phys. A}\ }\textbf {\bibinfo
  {volume} {32}},\ \bibinfo {pages} {1730006} (\bibinfo {year} {2017})},\
  \Eprint {http://arxiv.org/abs/1702.02430} {arXiv:1702.02430 [hep-ph]}
  \BibitemShut {NoStop}%
\bibitem [{\citenamefont {Penning}(2018)}]{Penning:2017tmb}%
  \BibitemOpen
  \bibfield  {author} {\bibinfo {author} {\bibfnamefont {B.}~\bibnamefont
  {Penning}},\ }\href {\doibase 10.1088/1361-6471/aabea7} {\bibfield  {journal}
  {\bibinfo  {journal} {J. Phys. G}\ }\textbf {\bibinfo {volume} {45}},\
  \bibinfo {pages} {063001} (\bibinfo {year} {2018})},\ \Eprint
  {http://arxiv.org/abs/1712.01391} {arXiv:1712.01391 [hep-ex]} \BibitemShut
  {NoStop}%
\bibitem [{\citenamefont {Chae}\ and\ \citenamefont
  {Perelstein}(2013)}]{Chae:2012bq}%
  \BibitemOpen
  \bibfield  {author} {\bibinfo {author} {\bibfnamefont {Y.~J.}\ \bibnamefont
  {Chae}}\ and\ \bibinfo {author} {\bibfnamefont {M.}~\bibnamefont
  {Perelstein}},\ }\href {\doibase 10.1007/JHEP05(2013)138} {\bibfield
  {journal} {\bibinfo  {journal} {JHEP}\ }\textbf {\bibinfo {volume} {05}},\
  \bibinfo {pages} {138} (\bibinfo {year} {2013})},\ \Eprint
  {http://arxiv.org/abs/1211.4008} {arXiv:1211.4008 [hep-ph]} \BibitemShut
  {NoStop}%
\bibitem [{\citenamefont {Habermehl}\ \emph {et~al.}(2020)\citenamefont
  {Habermehl}, \citenamefont {Berggren},\ and\ \citenamefont
  {List}}]{Habermehl:2020njb}%
  \BibitemOpen
  \bibfield  {author} {\bibinfo {author} {\bibfnamefont {M.}~\bibnamefont
  {Habermehl}}, \bibinfo {author} {\bibfnamefont {M.}~\bibnamefont {Berggren}},
  \ and\ \bibinfo {author} {\bibfnamefont {J.}~\bibnamefont {List}},\ }\href
  {\doibase 10.1103/PhysRevD.101.075053} {\bibfield  {journal} {\bibinfo
  {journal} {Phys. Rev. D}\ }\textbf {\bibinfo {volume} {101}},\ \bibinfo
  {pages} {075053} (\bibinfo {year} {2020})},\ \Eprint
  {http://arxiv.org/abs/2001.03011} {arXiv:2001.03011 [hep-ex]} \BibitemShut
  {NoStop}%
\bibitem [{\citenamefont {Kalinowski}\ \emph {et~al.}(2021)\citenamefont
  {Kalinowski}, \citenamefont {Kotlarski}, \citenamefont {Mekala},
  \citenamefont {Sopicki},\ and\ \citenamefont
  {Zarnecki}}]{Kalinowski:2021tyr}%
  \BibitemOpen
  \bibfield  {author} {\bibinfo {author} {\bibfnamefont {J.}~\bibnamefont
  {Kalinowski}}, \bibinfo {author} {\bibfnamefont {W.}~\bibnamefont
  {Kotlarski}}, \bibinfo {author} {\bibfnamefont {K.}~\bibnamefont {Mekala}},
  \bibinfo {author} {\bibfnamefont {P.}~\bibnamefont {Sopicki}}, \ and\
  \bibinfo {author} {\bibfnamefont {A.~F.}\ \bibnamefont {Zarnecki}},\
  }\href@noop {} {\  (\bibinfo {year} {2021})},\ \Eprint
  {http://arxiv.org/abs/2107.11194} {arXiv:2107.11194 [hep-ph]} \BibitemShut
  {NoStop}%
\bibitem [{\citenamefont {Kundu}\ \emph {et~al.}(2023)\citenamefont {Kundu},
  \citenamefont {Guha}, \citenamefont {Das},\ and\ \citenamefont
  {Dev}}]{Kundu:2021cmo}%
  \BibitemOpen
  \bibfield  {author} {\bibinfo {author} {\bibfnamefont {S.}~\bibnamefont
  {Kundu}}, \bibinfo {author} {\bibfnamefont {A.}~\bibnamefont {Guha}},
  \bibinfo {author} {\bibfnamefont {P.~K.}\ \bibnamefont {Das}}, \ and\
  \bibinfo {author} {\bibfnamefont {P.~S.~B.}\ \bibnamefont {Dev}},\ }\href
  {\doibase 10.1103/PhysRevD.107.015003} {\bibfield  {journal} {\bibinfo
  {journal} {Phys. Rev. D}\ }\textbf {\bibinfo {volume} {107}},\ \bibinfo
  {pages} {015003} (\bibinfo {year} {2023})},\ \Eprint
  {http://arxiv.org/abs/2110.06903} {arXiv:2110.06903 [hep-ph]} \BibitemShut
  {NoStop}%
\bibitem [{\citenamefont {Nowakowski}\ \emph {et~al.}(2005)\citenamefont
  {Nowakowski}, \citenamefont {Paschos},\ and\ \citenamefont
  {Rodriguez}}]{Nowakowski:2004cv}%
  \BibitemOpen
  \bibfield  {author} {\bibinfo {author} {\bibfnamefont {M.}~\bibnamefont
  {Nowakowski}}, \bibinfo {author} {\bibfnamefont {E.~A.}\ \bibnamefont
  {Paschos}}, \ and\ \bibinfo {author} {\bibfnamefont {J.~M.}\ \bibnamefont
  {Rodriguez}},\ }\href {\doibase 10.1088/0143-0807/26/4/001} {\bibfield
  {journal} {\bibinfo  {journal} {Eur. J. Phys.}\ }\textbf {\bibinfo {volume}
  {26}},\ \bibinfo {pages} {545} (\bibinfo {year} {2005})},\ \Eprint
  {http://arxiv.org/abs/physics/0402058} {arXiv:physics/0402058} \BibitemShut
  {NoStop}%
\bibitem [{\citenamefont {Giunti}\ and\ \citenamefont
  {Studenikin}(2015)}]{Giunti:2014ixa}%
  \BibitemOpen
  \bibfield  {author} {\bibinfo {author} {\bibfnamefont {C.}~\bibnamefont
  {Giunti}}\ and\ \bibinfo {author} {\bibfnamefont {A.}~\bibnamefont
  {Studenikin}},\ }\href {\doibase 10.1103/RevModPhys.87.531} {\bibfield
  {journal} {\bibinfo  {journal} {Rev. Mod. Phys.}\ }\textbf {\bibinfo {volume}
  {87}},\ \bibinfo {pages} {531} (\bibinfo {year} {2015})},\ \Eprint
  {http://arxiv.org/abs/1403.6344} {arXiv:1403.6344 [hep-ph]} \BibitemShut
  {NoStop}%
\bibitem [{\citenamefont {Belanger}\ \emph {et~al.}(2009)\citenamefont
  {Belanger}, \citenamefont {Boudjema}, \citenamefont {Pukhov},\ and\
  \citenamefont {Semenov}}]{Belanger:2008sj}%
  \BibitemOpen
  \bibfield  {author} {\bibinfo {author} {\bibfnamefont {G.}~\bibnamefont
  {Belanger}}, \bibinfo {author} {\bibfnamefont {F.}~\bibnamefont {Boudjema}},
  \bibinfo {author} {\bibfnamefont {A.}~\bibnamefont {Pukhov}}, \ and\ \bibinfo
  {author} {\bibfnamefont {A.}~\bibnamefont {Semenov}},\ }\href {\doibase
  10.1016/j.cpc.2008.11.019} {\bibfield  {journal} {\bibinfo  {journal}
  {Comput. Phys. Commun.}\ }\textbf {\bibinfo {volume} {180}},\ \bibinfo
  {pages} {747} (\bibinfo {year} {2009})},\ \Eprint
  {http://arxiv.org/abs/0803.2360} {arXiv:0803.2360 [hep-ph]} \BibitemShut
  {NoStop}%
\bibitem [{\citenamefont {Belanger}\ \emph {et~al.}(2021)\citenamefont
  {Belanger}, \citenamefont {Mjallal},\ and\ \citenamefont
  {Pukhov}}]{Belanger:2020gnr}%
  \BibitemOpen
  \bibfield  {author} {\bibinfo {author} {\bibfnamefont {G.}~\bibnamefont
  {Belanger}}, \bibinfo {author} {\bibfnamefont {A.}~\bibnamefont {Mjallal}}, \
  and\ \bibinfo {author} {\bibfnamefont {A.}~\bibnamefont {Pukhov}},\ }\href
  {\doibase 10.1140/epjc/s10052-021-09012-z} {\bibfield  {journal} {\bibinfo
  {journal} {Eur. Phys. J. C}\ }\textbf {\bibinfo {volume} {81}},\ \bibinfo
  {pages} {239} (\bibinfo {year} {2021})},\ \Eprint
  {http://arxiv.org/abs/2003.08621} {arXiv:2003.08621 [hep-ph]} \BibitemShut
  {NoStop}%
\bibitem [{\citenamefont {Aprile}\ \emph {et~al.}(2018)\citenamefont {Aprile}
  \emph {et~al.}}]{XENON:2018voc}%
  \BibitemOpen
  \bibfield  {author} {\bibinfo {author} {\bibfnamefont {E.}~\bibnamefont
  {Aprile}} \emph {et~al.} (\bibinfo {collaboration} {XENON}),\ }\href
  {\doibase 10.1103/PhysRevLett.121.111302} {\bibfield  {journal} {\bibinfo
  {journal} {Phys. Rev. Lett.}\ }\textbf {\bibinfo {volume} {121}},\ \bibinfo
  {pages} {111302} (\bibinfo {year} {2018})},\ \Eprint
  {http://arxiv.org/abs/1805.12562} {arXiv:1805.12562 [astro-ph.CO]}
  \BibitemShut {NoStop}%
\bibitem [{\citenamefont {Meng}\ \emph {et~al.}(2021)\citenamefont {Meng} \emph
  {et~al.}}]{PandaX-4T:2021bab}%
  \BibitemOpen
  \bibfield  {author} {\bibinfo {author} {\bibfnamefont {Y.}~\bibnamefont
  {Meng}} \emph {et~al.} (\bibinfo {collaboration} {PandaX-4T}),\ }\href@noop
  {} {\  (\bibinfo {year} {2021})},\ \Eprint {http://arxiv.org/abs/2107.13438}
  {arXiv:2107.13438 [hep-ex]} \BibitemShut {NoStop}%
\bibitem [{\citenamefont {Agnes}\ \emph {et~al.}(2018)\citenamefont {Agnes}
  \emph {et~al.}}]{DarkSide:2018bpj}%
  \BibitemOpen
  \bibfield  {author} {\bibinfo {author} {\bibfnamefont {P.}~\bibnamefont
  {Agnes}} \emph {et~al.} (\bibinfo {collaboration} {DarkSide}),\ }\href
  {\doibase 10.1103/PhysRevLett.121.081307} {\bibfield  {journal} {\bibinfo
  {journal} {Phys. Rev. Lett.}\ }\textbf {\bibinfo {volume} {121}},\ \bibinfo
  {pages} {081307} (\bibinfo {year} {2018})},\ \Eprint
  {http://arxiv.org/abs/1802.06994} {arXiv:1802.06994 [astro-ph.HE]}
  \BibitemShut {NoStop}%
\bibitem [{\citenamefont {Angloher}\ \emph {et~al.}(2016)\citenamefont
  {Angloher} \emph {et~al.}}]{CRESST:2015txj}%
  \BibitemOpen
  \bibfield  {author} {\bibinfo {author} {\bibfnamefont {G.}~\bibnamefont
  {Angloher}} \emph {et~al.} (\bibinfo {collaboration} {CRESST}),\ }\href
  {\doibase 10.1140/epjc/s10052-016-3877-3} {\bibfield  {journal} {\bibinfo
  {journal} {Eur. Phys. J. C}\ }\textbf {\bibinfo {volume} {76}},\ \bibinfo
  {pages} {25} (\bibinfo {year} {2016})},\ \Eprint
  {http://arxiv.org/abs/1509.01515} {arXiv:1509.01515 [astro-ph.CO]}
  \BibitemShut {NoStop}%
\bibitem [{\citenamefont {Agnese}\ \emph {et~al.}(2018)\citenamefont {Agnese}
  \emph {et~al.}}]{SuperCDMS:2017nns}%
  \BibitemOpen
  \bibfield  {author} {\bibinfo {author} {\bibfnamefont {R.}~\bibnamefont
  {Agnese}} \emph {et~al.} (\bibinfo {collaboration} {SuperCDMS}),\ }\href
  {\doibase 10.1103/PhysRevD.97.022002} {\bibfield  {journal} {\bibinfo
  {journal} {Phys. Rev. D}\ }\textbf {\bibinfo {volume} {97}},\ \bibinfo
  {pages} {022002} (\bibinfo {year} {2018})},\ \Eprint
  {http://arxiv.org/abs/1707.01632} {arXiv:1707.01632 [astro-ph.CO]}
  \BibitemShut {NoStop}%
\bibitem [{\citenamefont {Billard}\ \emph
  {et~al.}(2014{\natexlab{b}})\citenamefont {Billard} \emph
  {et~al.}}]{Billard:2013cxa}%
  \BibitemOpen
  \bibfield  {author} {\bibinfo {author} {\bibfnamefont {J.}~\bibnamefont
  {Billard}} \emph {et~al.},\ }\href {\doibase 10.1088/1748-0221/9/01/P01013}
  {\bibfield  {journal} {\bibinfo  {journal} {JINST}\ }\textbf {\bibinfo
  {volume} {9}},\ \bibinfo {pages} {P01013} (\bibinfo {year}
  {2014}{\natexlab{b}})},\ \Eprint {http://arxiv.org/abs/1305.2360}
  {arXiv:1305.2360 [astro-ph.IM]} \BibitemShut {NoStop}%
\bibitem [{\citenamefont {Ruppin}\ \emph {et~al.}(2014)\citenamefont {Ruppin},
  \citenamefont {Billard}, \citenamefont {Figueroa-Feliciano},\ and\
  \citenamefont {Strigari}}]{Ruppin:2014bra}%
  \BibitemOpen
  \bibfield  {author} {\bibinfo {author} {\bibfnamefont {F.}~\bibnamefont
  {Ruppin}}, \bibinfo {author} {\bibfnamefont {J.}~\bibnamefont {Billard}},
  \bibinfo {author} {\bibfnamefont {E.}~\bibnamefont {Figueroa-Feliciano}}, \
  and\ \bibinfo {author} {\bibfnamefont {L.}~\bibnamefont {Strigari}},\ }\href
  {\doibase 10.1103/PhysRevD.90.083510} {\bibfield  {journal} {\bibinfo
  {journal} {Phys. Rev. D}\ }\textbf {\bibinfo {volume} {90}},\ \bibinfo
  {pages} {083510} (\bibinfo {year} {2014})},\ \Eprint
  {http://arxiv.org/abs/1408.3581} {arXiv:1408.3581 [hep-ph]} \BibitemShut
  {NoStop}%
\bibitem [{\citenamefont {Davis}(2015)}]{Davis:2014ama}%
  \BibitemOpen
  \bibfield  {author} {\bibinfo {author} {\bibfnamefont {J.~H.}\ \bibnamefont
  {Davis}},\ }\href {\doibase 10.1088/1475-7516/2015/03/012} {\bibfield
  {journal} {\bibinfo  {journal} {JCAP}\ }\textbf {\bibinfo {volume} {03}},\
  \bibinfo {pages} {012} (\bibinfo {year} {2015})},\ \Eprint
  {http://arxiv.org/abs/1412.1475} {arXiv:1412.1475 [hep-ph]} \BibitemShut
  {NoStop}%
\bibitem [{\citenamefont {Sassi}\ \emph {et~al.}(2021)\citenamefont {Sassi},
  \citenamefont {Dinmohammadi}, \citenamefont {Heikinheimo}, \citenamefont
  {Mirabolfathi}, \citenamefont {Nordlund}, \citenamefont {Safari},\ and\
  \citenamefont {Tuominen}}]{Sassi:2021umf}%
  \BibitemOpen
  \bibfield  {author} {\bibinfo {author} {\bibfnamefont {S.}~\bibnamefont
  {Sassi}}, \bibinfo {author} {\bibfnamefont {A.}~\bibnamefont {Dinmohammadi}},
  \bibinfo {author} {\bibfnamefont {M.}~\bibnamefont {Heikinheimo}}, \bibinfo
  {author} {\bibfnamefont {N.}~\bibnamefont {Mirabolfathi}}, \bibinfo {author}
  {\bibfnamefont {K.}~\bibnamefont {Nordlund}}, \bibinfo {author}
  {\bibfnamefont {H.}~\bibnamefont {Safari}}, \ and\ \bibinfo {author}
  {\bibfnamefont {K.}~\bibnamefont {Tuominen}},\ }\href {\doibase
  10.1103/PhysRevD.104.063037} {\bibfield  {journal} {\bibinfo  {journal}
  {Phys. Rev. D}\ }\textbf {\bibinfo {volume} {104}},\ \bibinfo {pages}
  {063037} (\bibinfo {year} {2021})},\ \Eprint
  {http://arxiv.org/abs/2103.08511} {arXiv:2103.08511 [hep-ph]} \BibitemShut
  {NoStop}%
\bibitem [{\citenamefont {O'Hare}(2021{\natexlab{b}})}]{OHare:2021utq}%
  \BibitemOpen
  \bibfield  {author} {\bibinfo {author} {\bibfnamefont {C.~A.~J.}\
  \bibnamefont {O'Hare}},\ }\href@noop {} {\  (\bibinfo {year}
  {2021}{\natexlab{b}})},\ \Eprint {http://arxiv.org/abs/2109.03116}
  {arXiv:2109.03116 [hep-ph]} \BibitemShut {NoStop}%
\bibitem [{\citenamefont {Alloul}\ \emph {et~al.}(2014)\citenamefont {Alloul},
  \citenamefont {Christensen}, \citenamefont {Degrande}, \citenamefont {Duhr},\
  and\ \citenamefont {Fuks}}]{Alloul:2013bka}%
  \BibitemOpen
  \bibfield  {author} {\bibinfo {author} {\bibfnamefont {A.}~\bibnamefont
  {Alloul}}, \bibinfo {author} {\bibfnamefont {N.~D.}\ \bibnamefont
  {Christensen}}, \bibinfo {author} {\bibfnamefont {C.}~\bibnamefont
  {Degrande}}, \bibinfo {author} {\bibfnamefont {C.}~\bibnamefont {Duhr}}, \
  and\ \bibinfo {author} {\bibfnamefont {B.}~\bibnamefont {Fuks}},\ }\href
  {\doibase 10.1016/j.cpc.2014.04.012} {\bibfield  {journal} {\bibinfo
  {journal} {Comput. Phys. Commun.}\ }\textbf {\bibinfo {volume} {185}},\
  \bibinfo {pages} {2250} (\bibinfo {year} {2014})},\ \Eprint
  {http://arxiv.org/abs/1310.1921} {arXiv:1310.1921 [hep-ph]} \BibitemShut
  {NoStop}%
\bibitem [{\citenamefont {Kilian}\ \emph {et~al.}(2011)\citenamefont {Kilian},
  \citenamefont {Ohl},\ and\ \citenamefont {Reuter}}]{Kilian:2007gr}%
  \BibitemOpen
  \bibfield  {author} {\bibinfo {author} {\bibfnamefont {W.}~\bibnamefont
  {Kilian}}, \bibinfo {author} {\bibfnamefont {T.}~\bibnamefont {Ohl}}, \ and\
  \bibinfo {author} {\bibfnamefont {J.}~\bibnamefont {Reuter}},\ }\href
  {\doibase 10.1140/epjc/s10052-011-1742-y} {\bibfield  {journal} {\bibinfo
  {journal} {Eur. Phys. J. C}\ }\textbf {\bibinfo {volume} {71}},\ \bibinfo
  {pages} {1742} (\bibinfo {year} {2011})},\ \Eprint
  {http://arxiv.org/abs/0708.4233} {arXiv:0708.4233 [hep-ph]} \BibitemShut
  {NoStop}%
\bibitem [{\citenamefont {Kalinowski}\ \emph {et~al.}(2020)\citenamefont
  {Kalinowski}, \citenamefont {Kotlarski}, \citenamefont {Sopicki},\ and\
  \citenamefont {Zarnecki}}]{Kalinowski:2020lhp}%
  \BibitemOpen
  \bibfield  {author} {\bibinfo {author} {\bibfnamefont {J.}~\bibnamefont
  {Kalinowski}}, \bibinfo {author} {\bibfnamefont {W.}~\bibnamefont
  {Kotlarski}}, \bibinfo {author} {\bibfnamefont {P.}~\bibnamefont {Sopicki}},
  \ and\ \bibinfo {author} {\bibfnamefont {A.~F.}\ \bibnamefont {Zarnecki}},\
  }\href {\doibase 10.1140/epjc/s10052-020-8149-6} {\bibfield  {journal}
  {\bibinfo  {journal} {Eur. Phys. J. C}\ }\textbf {\bibinfo {volume} {80}},\
  \bibinfo {pages} {634} (\bibinfo {year} {2020})},\ \Eprint
  {http://arxiv.org/abs/2004.14486} {arXiv:2004.14486 [hep-ph]} \BibitemShut
  {NoStop}%
\bibitem [{\citenamefont {de~Favereau}\ \emph {et~al.}(2014)\citenamefont
  {de~Favereau}, \citenamefont {Delaere}, \citenamefont {Demin}, \citenamefont
  {Giammanco}, \citenamefont {Lema\^\i{}tre}, \citenamefont {Mertens},\ and\
  \citenamefont {Selvaggi}}]{deFavereau:2013fsa}%
  \BibitemOpen
  \bibfield  {author} {\bibinfo {author} {\bibfnamefont {J.}~\bibnamefont
  {de~Favereau}}, \bibinfo {author} {\bibfnamefont {C.}~\bibnamefont
  {Delaere}}, \bibinfo {author} {\bibfnamefont {P.}~\bibnamefont {Demin}},
  \bibinfo {author} {\bibfnamefont {A.}~\bibnamefont {Giammanco}}, \bibinfo
  {author} {\bibfnamefont {V.}~\bibnamefont {Lema\^\i{}tre}}, \bibinfo {author}
  {\bibfnamefont {A.}~\bibnamefont {Mertens}}, \ and\ \bibinfo {author}
  {\bibfnamefont {M.}~\bibnamefont {Selvaggi}} (\bibinfo {collaboration}
  {DELPHES 3}),\ }\href {\doibase 10.1007/JHEP02(2014)057} {\bibfield
  {journal} {\bibinfo  {journal} {JHEP}\ }\textbf {\bibinfo {volume} {02}},\
  \bibinfo {pages} {057} (\bibinfo {year} {2014})},\ \Eprint
  {http://arxiv.org/abs/1307.6346} {arXiv:1307.6346 [hep-ex]} \BibitemShut
  {NoStop}%
\bibitem [{\citenamefont {Barklow}\ \emph {et~al.}(2015)\citenamefont
  {Barklow}, \citenamefont {Brau}, \citenamefont {Fujii}, \citenamefont {Gao},
  \citenamefont {List}, \citenamefont {Walker},\ and\ \citenamefont
  {Yokoya}}]{Barklow:2015tja}%
  \BibitemOpen
  \bibfield  {author} {\bibinfo {author} {\bibfnamefont {T.}~\bibnamefont
  {Barklow}}, \bibinfo {author} {\bibfnamefont {J.}~\bibnamefont {Brau}},
  \bibinfo {author} {\bibfnamefont {K.}~\bibnamefont {Fujii}}, \bibinfo
  {author} {\bibfnamefont {J.}~\bibnamefont {Gao}}, \bibinfo {author}
  {\bibfnamefont {J.}~\bibnamefont {List}}, \bibinfo {author} {\bibfnamefont
  {N.}~\bibnamefont {Walker}}, \ and\ \bibinfo {author} {\bibfnamefont
  {K.}~\bibnamefont {Yokoya}},\ }\href@noop {} {\  (\bibinfo {year} {2015})},\
  \Eprint {http://arxiv.org/abs/1506.07830} {arXiv:1506.07830 [hep-ex]}
  \BibitemShut {NoStop}%
\bibitem [{\citenamefont {Brun}\ and\ \citenamefont
  {Rademakers}(1997)}]{BRUN199781}%
  \BibitemOpen
  \bibfield  {author} {\bibinfo {author} {\bibfnamefont {R.}~\bibnamefont
  {Brun}}\ and\ \bibinfo {author} {\bibfnamefont {F.}~\bibnamefont
  {Rademakers}},\ }\href {\doibase
  https://doi.org/10.1016/S0168-9002(97)00048-X} {\bibfield  {journal}
  {\bibinfo  {journal} {Nuclear Instruments and Methods in Physics Research
  Section A: Accelerators, Spectrometers, Detectors and Associated Equipment}\
  }\textbf {\bibinfo {volume} {389}},\ \bibinfo {pages} {81} (\bibinfo {year}
  {1997})},\ \bibinfo {note} {new Computing Techniques in Physics Research
  V}\BibitemShut {NoStop}%
\bibitem [{\citenamefont {Habermehl}(2018)}]{Habermehl:2018yul}%
  \BibitemOpen
  \bibfield  {author} {\bibinfo {author} {\bibfnamefont {M.}~\bibnamefont
  {Habermehl}},\ }\emph {\bibinfo {title} {{Dark Matter at the International
  Linear Collider}}},\ \href {\doibase 10.3204/PUBDB-2018-05723} {Ph.D.
  thesis},\ \bibinfo  {school} {Hamburg U.}, \bibinfo {address} {Hamburg}
  (\bibinfo {year} {2018})\BibitemShut {NoStop}%
\bibitem [{\citenamefont {Abramowicz}\ \emph {et~al.}(2010)\citenamefont
  {Abramowicz} \emph {et~al.}}]{Abramowicz:2010bg}%
  \BibitemOpen
  \bibfield  {author} {\bibinfo {author} {\bibfnamefont {H.}~\bibnamefont
  {Abramowicz}} \emph {et~al.},\ }\href {\doibase
  10.1088/1748-0221/5/12/P12002} {\bibfield  {journal} {\bibinfo  {journal}
  {JINST}\ }\textbf {\bibinfo {volume} {5}},\ \bibinfo {pages} {P12002}
  (\bibinfo {year} {2010})},\ \Eprint {http://arxiv.org/abs/1009.2433}
  {arXiv:1009.2433 [physics.ins-det]} \BibitemShut {NoStop}%
\bibitem [{\citenamefont {Blaising}\ \emph {et~al.}(2021)\citenamefont
  {Blaising}, \citenamefont {Roloff}, \citenamefont {Sailer},\ and\
  \citenamefont {Schnoor}}]{Blaising:2021vhh}%
  \BibitemOpen
  \bibfield  {author} {\bibinfo {author} {\bibfnamefont {J.-J.}\ \bibnamefont
  {Blaising}}, \bibinfo {author} {\bibfnamefont {P.}~\bibnamefont {Roloff}},
  \bibinfo {author} {\bibfnamefont {A.}~\bibnamefont {Sailer}}, \ and\ \bibinfo
  {author} {\bibfnamefont {U.}~\bibnamefont {Schnoor}} (\bibinfo
  {collaboration} {CLICdp}),\ }\href@noop {} {\  (\bibinfo {year} {2021})},\
  \Eprint {http://arxiv.org/abs/2103.06006} {arXiv:2103.06006 [hep-ex]}
  \BibitemShut {NoStop}%
\end{thebibliography}%

\end{document}